\documentclass[graybox]{svmult}

\usepackage{type1cm}        
\usepackage{makeidx}         
\usepackage{graphicx}        
\usepackage{multicol}        
\usepackage[bottom]{footmisc}
\usepackage{lineno}
\usepackage{booktabs}

\usepackage{newtxtext}       %
\usepackage[varvw]{newtxmath}       

\makeindex             

\begin{document}

\title*{Neural network analysis of X-ray polarimeter data}
\author{A.L.Peirson}
\institute{Abel Lawrence Peirson \at Kavli Institute for Particle Astrophysics and Cosmology,
Stanford University,
Stanford, CA, 94305 \email{alpv95@stanford.edu}}
%
%
\maketitle

\abstract*{This chapter presents deep neural network based methods for enhancing the sensitivity of X-ray telescopic observations with imaging polarimeters. Deep neural networks can be used to determine photoelectron emission directions, photon absorptions points, and photon energies from 2D photoelectron track images, with estimates for both the statistical and model uncertainties. Deep neural network predictive uncertainties can be incorporated into a weighted maximum likelihood to estimate source polarization parameters. Events converting outside of the fiducial gas volume, whose tracks have little polarization sensitivity, complicate polarization estimation. Deep neural network based classifiers can be used to select against these events to improve energy resolution and polarization sensitivity. The performance of deep neural network methods is compared against standard data analysis methods, revealing a $< 0.75\times$ improvement in minimum detectable polarization for IXPE-specific simulations. Potential future developments and improvements to these methods are discussed.}

\abstract{This chapter presents deep neural network based methods for enhancing the sensitivity of X-ray telescopic observations with imaging polarimeters. Deep neural networks can be used to determine photoelectron emission directions, photon absorptions points, and photon energies from 2D photoelectron track images, with estimates for both the statistical and model uncertainties. Deep neural network predictive uncertainties can be incorporated into a weighted maximum likelihood to estimate source polarization parameters. Events converting outside of the fiducial gas volume, whose tracks have little polarization sensitivity, complicate polarization estimation. Deep neural network based classifiers can be used to select against these events to improve energy resolution and polarization sensitivity. The performance of deep neural network methods is compared against standard data analysis methods, revealing a $< 0.75\times$ improvement in minimum detectable polarization for IXPE-specific simulations. Potential future developments and improvements to these methods are discussed.
}
\tableofcontents
\newpage

\section{Introduction}
\label{sec:1}
Measuring X-ray polarization has been a major goal in astrophysics for the last 40 years. X-ray polarization measurements offer rich opportunities to probe the magnetic field topology and emission physics of high energy astrophysical sources, such as accreting black holes and astrophysical jets \cite{krawczynski_using_2019, weisskopf_overview_2018}. 
The recent development of photoelectron tracking detectors \cite{bellazzini_novel_2003} has greatly improved the prospects of doing so. The gas pixel detector (GPD) \cite{bellazzini_sealed_2007} has brought soft X-ray polarimetry (1-10 keV) to the PolarLight CubeSat test \cite{feng_x-ray_2020}, the scheduled NASA IXPE mission \cite{sgro_imaging_2019}, and the potential Chinese mission, eXTP \cite{zhang_extp_2017}. 

\begin{figure}[b]
\sidecaption
\includegraphics[scale=.75]{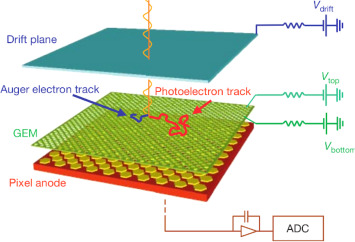}
\caption{Design of the GPD. An incoming X-ray photon interacts with a gas molecule to produce a photoelectron between the entrance window and the gas electron multiplier (GEM). Secondary charges created along the photoelectron's path are multiplied by the GEM and read out at the anode plane. This forms a 2D photoelectron track image on hexagonal pixel grid; fig.~\ref{fig:tracks} gives some examples. Figure from \cite{baldini_design_2021}.}
\label{fig:gpd}       
\end{figure}

\begin{figure}[t]
\includegraphics[scale=.45]{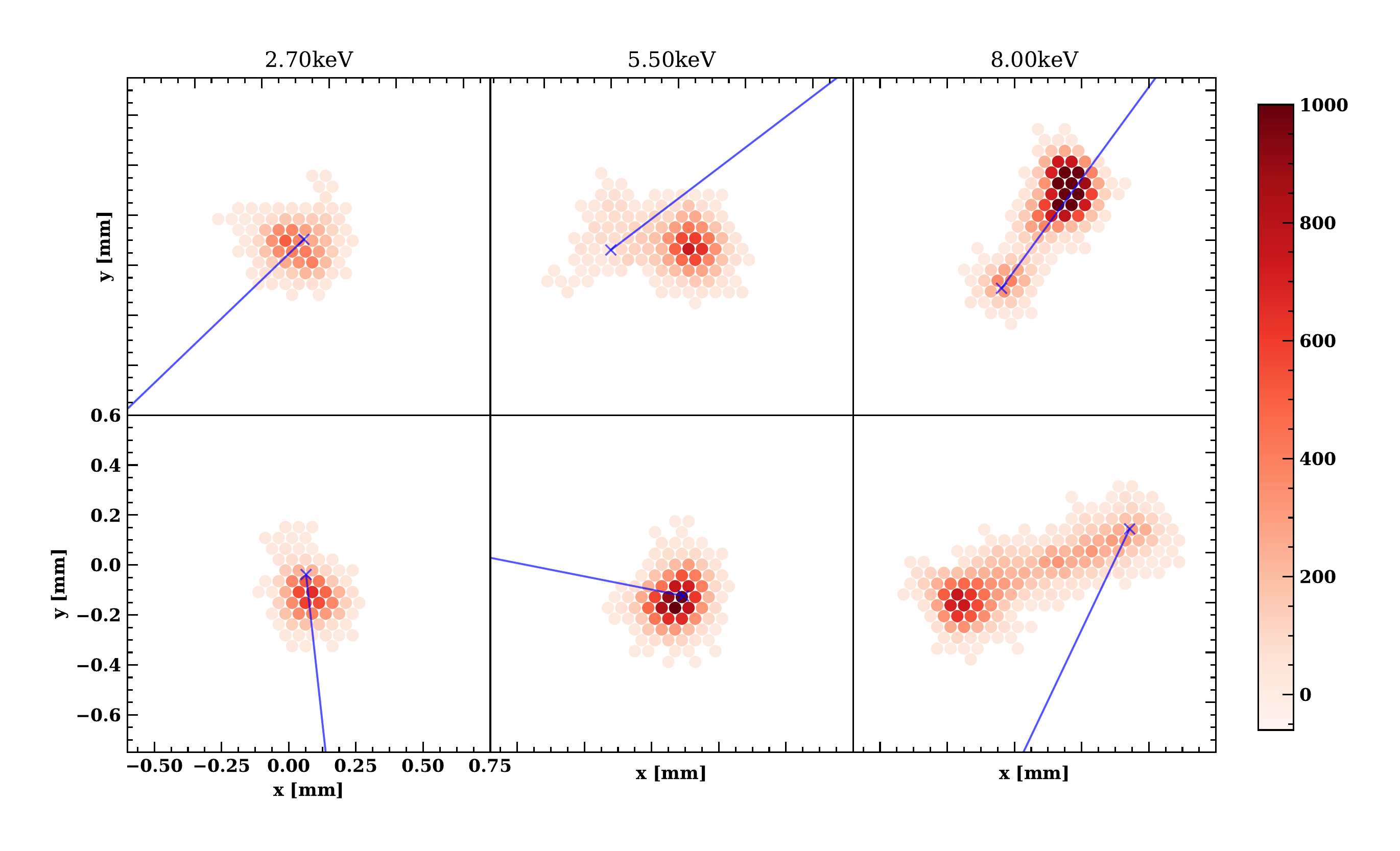}
\caption{A selection of measured photoelectron track events from IXPE GPDs. Pixel color density represents charge deposited, blue crosses show photon x-y absorption point and blue lines the initial photoelectron direction. Pixels are thresholded at 10 counts. Track morphologies vary widely and depend strongly on energy.}
\label{fig:tracks}       
\end{figure}

X-ray polarization telescopes using GPDs directly image electron tracks formed from photoelectrons scattered by incoming X-ray photons. This technique has the capability to build up an image of extended sources (e.g., supernova remnants or pulsar wind nebulae). Fig.~\ref{fig:gpd} shows a schematic of the GPD and fig.~\ref{fig:tracks} gives example photoelectron tracks at various photon energies, measured by IXPE's GPDs. 
GPD sensitivity is limited by the track analysis algorithm used to recover source polarization, spatial structure, and energy, given a measured set of electron track images.

\begin{figure}[t]
\sidecaption
\includegraphics[scale=.85]{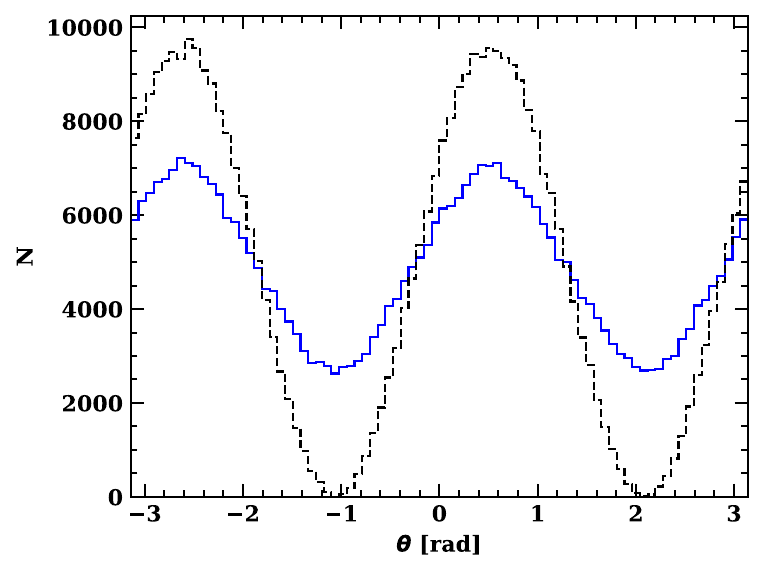}
\caption{Photoelectron angles for a $100\%$ polarized ($p_0 = 1$) 6.4keV simulated line source. Black gives the true photoelectron angles, blue gives the recovered photoelectron angles for the standard moment analysis.}
\label{fig:mu}       
\end{figure}

In the $1-15$ keV range, the cross-section for photoelectron emission is proportional to cos$^2(\theta - \theta_0)$, where $\theta_0$ is the normal incidence X-ray's electric vector position angle (EVPA) and $\theta$ the azimuthal emission direction of the photoelectron; see the previous chapter. 
Specifically, the emission angles $\theta$ follow the distribution 
\begin{equation}
    \theta \sim \frac{1}{2\pi} \big(1 + p_0\cos[2(\theta - \theta_0)] \big).
    \label{eqn:prob}
\end{equation}
By measuring a large number of individual photoelectron emission angles $\{\hat{\theta}_i\}_{i=1}^N$, one can recover the above distribution to extract the source polarization parameters: polarization fraction ($0 \leq p_0 \leq 1$) and EVPA ($-\pi/2 \leq \theta_0 < \pi/2$). In practice, the recovery of photoelectron emission angles from track images is imperfect. Track images are noisy due to Coulomb scattering and diffusion, and, especially for low energies, are often barely resolved. For example, in fig.~\ref{fig:mu} the distribution of recovered emission angles using IXPE's classical track reconstruction is significantly blurred compared to the true distribution.

The best classical track reconstruction method for GPDs is a moment analysis described by \cite{bellazzini_novel_2003} and in the previous chapter. Impressive accuracies for the emission angle and photon absorption point are achieved from a simple weighted combination of track moments. Photon energy estimates are proportional to the total collected GPD charge for a track. The track ellipticity is quantified to provide a rough proxy for track reconstruction quality. High ellipticity tracks typically have more accurate angle estimates. However, simple moments cannot capture all image information, especially for long high energy tracks, and so a more sophisticated image analysis scheme can lead to improved track emission angle, absorption point, and energy recovery.

Once the emission angles have been reconstructed, $\{\hat{\theta}_i\}_{i=1}^N$,
classical polarization estimation uses a maximum likelihood estimator (MLE) or direct curve fit to calculate $(\hat{p}_0, \hat{\theta}_0)$. This assumes individual tracks contribute equally to the final polarization estimate. In fact, photoelectron tracks are very morphologically diverse, even for the same photon energy (cf.~fig.~\ref{fig:tracks}), and so emission angle estimates are highly heteroskedastic: some emission angles are much better estimated than others. 
This is especially true for tracks of different photon energies, important in the case of broadband polarization estimates. Assuming an equal contribution from all estimated emission angles results in sub-optimal polarization recovery. 
To ameliorate this, IXPE's standard analysis performs an event cut removing tracks in the bottom $20\%$ of estimated ellipticities; this does provide a marginal improvement in the recovered polarization signal. However, a detailed analysis with proper estimation of emission angle uncertainties and their inclusion in the likelihood function can substantially improve the signal-to-noise ratio of recovered polarization.


Neural networks are supervised learning models that are the state of the art for classification and regression problems in many fields, especially computer vision \cite{mahony_deep_2020}. They are highly non-linear models trained on large datasets of labelled input-output example pairs. Neural networks are often criticized as black box models with no output interpretability. However, recent work in uncertainty quantification for neural networks \cite{abdar_review_2021} has demonstrated these models are capable of trustworthy uncertainty estimates on their predictions. Neural networks' success in dealing with image data and their advances in uncertainty quantification make them a perfect candidate for X-ray polarimetry track analysis \cite{kitaguchi_convolutional_2019, peirson_deep_2021}.

This chapter outlines neural networks approaches to track analysis algorithms, covering both track reconstruction and polarization estimation. 
The methods discussed in this chapter were developed primarily for GPDs, but should be applicable to all imaging X-ray polarimetry technologies, such as the time projection chambers proposed for the former GEMS/Praxys mission \cite{iwakiri_performance_2016}. 


\subsection{How this chapter is organized}
\begin{itemize}
\item \S\ref{sec:2} contains an overview of the data analysis process necessary to go from telescope observations to polarization measurements and energy spectra, including some of the standard pipeline, covered in more detail in the previous chapter.
\item \S\ref{sec:3} gives a brief overview of neural network models and the types of deep learning important for X-ray polarimetry, with a discussion of uncertainty quantification for neural networks.
\item \S\ref{sec:4} describes an application of neural networks to track reconstruction: extracting emission angles, absorption points and photon energies from GPD track images. Comparison of neural network methods to the standard data analysis.
A neural network approach to identifying and removing events outside of the main detector.
\item \S\ref{sec:5} illustrates the use of neural network results for polarization estimation: using neural network predicted uncertainties on emission angles to form a proper maximum likelihood estimator of the polarization parameters.
Comparison of neural network methods to the standard data analysis.
Discussion of the impacts of events outside of the main detector gas.
\item \S\ref{sec:6} contains the conclusion, limitations, and future directions in the field.
\end{itemize}

\newpage
\section{Imaging X-ray polarimetry}
\label{sec:2}
Measuring the X-ray polarization of a source using an imaging X-ray polarimeter is, from a data analysis perspective, a two step process. 
\begin{enumerate}
    \item \textit{Track reconstruction}. The telescope is pointed at the source and collects photoelectron track images, see fig.~\ref{fig:tracks}. Individual track images are processed to extract all of the relevant features: emission angles, absorption points and photon energies. \\
    
    \item \textit{Polarization estimation}. The extracted emission angles from many individual photoelectron tracks are combined to estimate the source polarization parameters.
\end{enumerate}
For spatially extended, time-varying or spectrally varying sources, the extracted emission angles could be grouped based on absorption points, arrival time or photon energy, i.e. in spatial, time and/or energy bins. Polarization estimation from these grouped emission angles follows the same procedure.

This chapter describes classical approaches for both steps, a prerequisite for the neural network approach. In classical polarization analysis the two steps are entirely disconnected. \S\ref{sec:5} will show how these two steps can be more closely connected using neural network uncertainty estimates.

\subsection{Track reconstruction}
\label{sec:trckrec}
\begin{figure}[t]
\sidecaption
\includegraphics[scale=.75]{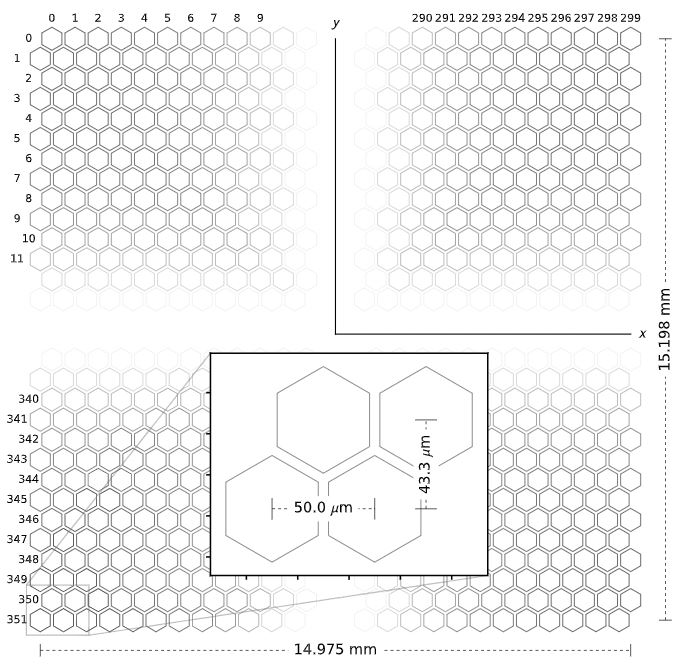}
\caption{Layout of IXPE GPD readout pixel array. There are approximately 90000 pixels. More pixels allow for better track reconstruction. Figure from \cite{bellazzini_novel_2003}.}
\label{fig:pix}       
\end{figure}

Photoelectron track images contain all the extractable information in imaging X-ray polarimeters. More specifically, the various sensitivities of an imaging X-ray polarimeter can be attributed to a few photoelectron track features that can be extracted from individual track images:
\begin{itemize}
    \item \textit{Polarization sensitivity:} photoelectron emission angle $\theta$, \S\ref{sec:emit}.
    \item \textit{Spatial sensitivity:} photon absorption point $(x,y)$, \S\ref{sec:absp}.
    \item \textit{Spectral sensitivity:} photon energy $E$, \S\ref{sec:E}.
\end{itemize}
Tracks are imaged by charge deposition onto an array of pixels. For example, IXPE's GPD pixel array is a $\sim$ 15mm x 15mm square tiled with hexagonal pixels on a $50\mu$m pitch \cite{baldini_design_2021}; see fig.~\ref{fig:pix}. The charge deposited in each pixel is measured in integer counts, typically ranging from 0 to 1000+, fig.~\ref{fig:tracks}. Track images are identified on the pixel array by clustering analysis and the region of interest is cropped. Each track image comes as a list of $(x,y,c)$ tuples: the pixel coordinates $x,y$ and charge deposited $c$. 

There are a number of instrument specific settings that affect track morphology. For example, a pixelwise charge threshold is usually applied to individual track images to reduce small noisy pixel clusters. The GPD gas pressure and composition can also be varied; lower pressure allows for longer tracks that are easier to reconstruct but reduces the detector quantum efficiency. The methods presented in this chapter generalize to all imaging polarimeter settings, details on optimizing specific settings can be found in chapters XY. The examples shown here assume IXPE-specific GPD settings.

\subsubsection{Emission angle reconstruction}
\label{sec:emit}
Source polarization is encoded in the photoelectron emission angles $\theta$. These are the image plane projected initial directions of the photoelectron at the interaction point. Track images capture the entire track of the photoelectron in the image plane, so in principle it should be possible to recover the initial direction.
In practice, track images are relatively noisy representations of the photoelectron path. Charge diffusion means the further away from the pixel array a photoelectron is emitted, the less well defined its track image will be.

Track images are typically asymmetric and follow a similar pattern: the photoelectron starts at the absorption point and moves along its initial direction. Sometimes an Auger electron is also released at the absorption point. As the photoelectron moves away from the absorption point, it is Coulomb scattered away from the initial direction. As the photoelectron slows down, it ionizes more gas molecules, depositing more charge in the pixel array and culminates in a final Bragg peak. Low energy photoelectrons travel less far from the absorption point and leave nearly circularly symmetric track images. 

The scattering of the photoelectron from its initial direction and short, symmetric photoelectron tracks make emission angle reconstruction challenging. The moment analysis described in \cite{bellazzini_novel_2003} uses the known structure of photoelectron tracks to improve their emission angle estimates. They initially calculate the principal axis of the track image by maximizing the second charge moment about the track barycenter. This provides a simple initial estimate of the emission angle, but is skewed by the Bragg peak -- the dense final part of the track that is usually not correlated to the initial direction because of Coulomb scattering. To correct for the Bragg peak bias, the location of the Bragg peak is calculated by observing the sign of the third charge moment about the track barycenter. This identifies which side of the track contains the Bragg peak. The absorption point is estimated as being a multiple of the second moment away from the track barycenter along the principal axis, away from the Bragg peak. Finally, the emission angle is calculated by using the second moment about the absorption point, using only pixels in the initial half of the track (those not on the Bragg peak side). See the previous chapter for more details on the moment analysis.


\subsubsection{Absorption point reconstruction}
\label{sec:absp}
Photon absorption points $(x,y)$ on the image plane along with the known point spread function (PSF) of the telescope optics allow for the spatial resolution of extended sources. Better absorption estimates yield better spatial resolution. A simple estimator for the absorption point of an individual photon from its track image is the track barycenter, but this can be strongly biased by the photoelectron Bragg peak \S\ref{sec:emit}. In moment analysis \cite{bellazzini_novel_2003}, the absorption point is estimated jointly with the emission angle by identifying and excluding the Bragg peak.
For low energy symmetric tracks where it is impossible to separate the Bragg peak, the track barycenter is used as the standard absorption point estimator.


\subsubsection{Energy reconstruction}
\label{sec:E}
An X-ray photon's energy is proportional to the average charge deposited in the detector. Since the charge deposited is subject to statistical fluctuations, the energy resolution is limited and this limit depends on the effective detector Fano factor. Classical polarimetry uses a simple linear model calibrated on real detector track images to reconstruct the photon energy. Thus, the total track charge is assumed proportional to the photon energy. This method approaches but does not meet the theoretical best energy resolution.

\begin{figure}[t]
\includegraphics[scale=.45]{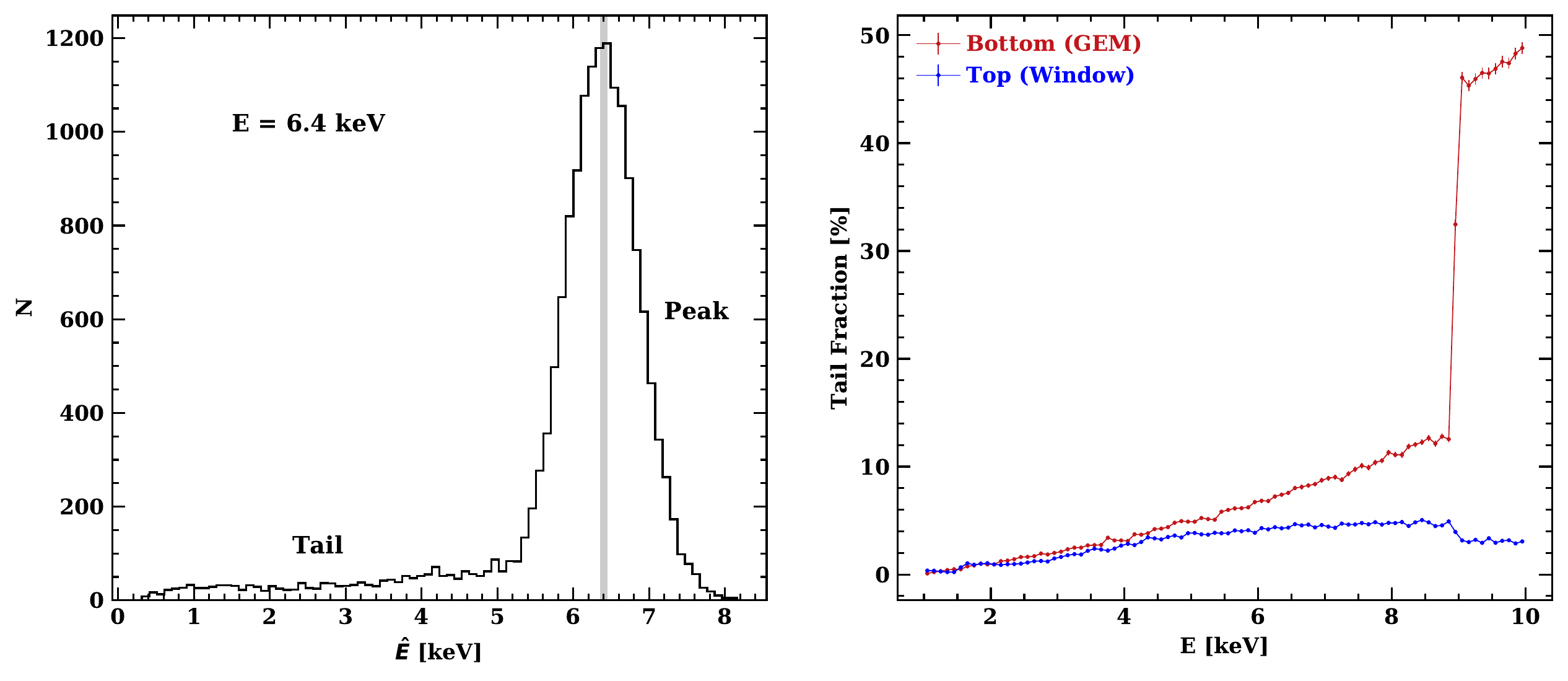}
\caption{\textit{Left:} Recovered energy histogram for a 6.4keV line source. A simple linear function of total charge deposited is used here for recovered energy, as in the standard moment analysis. The long low energy tail is produced by events converting in the window or GEM. \textit{Right:} Fraction of events that are `tails' as a function of energy. Red and blue traces show the Be window and GEM conversion respectively. The jump in GEM conversions at 8.9\,keV Cu edge is prominent.}
\label{fig:tails}       
\end{figure}

\subsubsection{Events converting outside of the gas volume}
\label{sec:tail}
So far, the track images considered are formed when photons interact with the detector gas. However, incoming X-ray photons can also convert in detector components just outside of the main gas volume, with electrons penetrating the gas triggering the GPD and producing photoelecton tracks. For IXPE's GPDs, these external interactions occur in the thin beryllium entrance window at the top of the GPD and in the gas electron multiplier copper material at the bottom, fig.~\ref{fig:gpd}. The track images formed by these external interactions have different morphological properties compared to those in the detector gas. Notably, they have less total charge deposited for the same photon energy, because some of the photoelectron energy can be lost in the solid window or GEM material. In addition they tend to have a denser `core' since the photoelectrons surviving to the gas volume tend to have trajectories with a large vertical component. The left panel of fig.~\ref{fig:tails} shows the linearly reconstructed energy (\S\ref{sec:E}) for a 6.4keV line source. External events form a low energy tail on the predicted energy histogram -- external events are described here tail events, producing tail tracks, while events converting in the detector gas are called peak events. The right panel gives the population fraction of tail events as a function of photon energy. Tail tracks are mostly subdominant, but with a linearly increasing fraction as photon energy increases. At energies above 8.9 keV, the Copper K absorption edge causes a large increase in tail events in the GEM, increasing the tail track fraction.

\begin{figure}[t]
\sidecaption
\includegraphics[scale=.65]{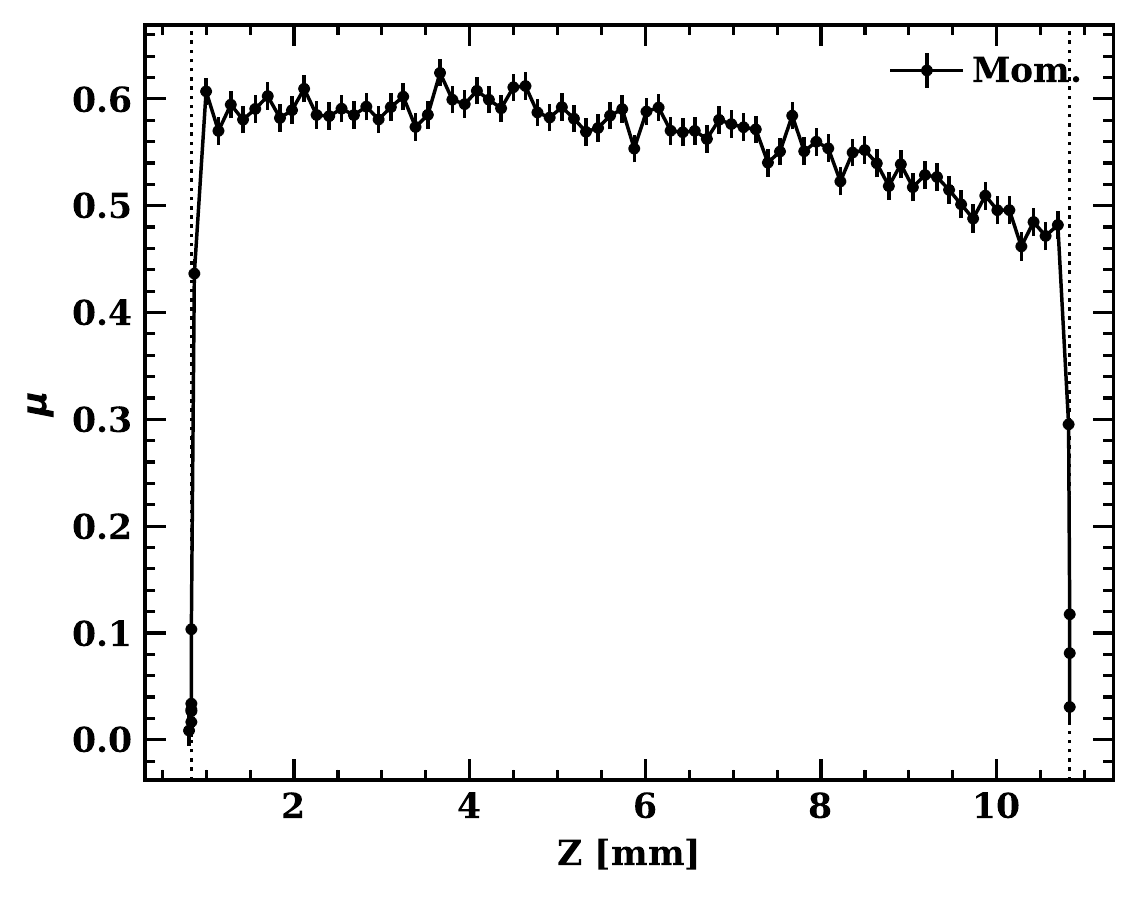}
\caption{Polarization sensitivity, measured by the modulation factor $\mu$ (\S\ref{sec:polest}), as a function of absorption point height in IXPE's GPD. Dotted lines denote the edges of the GPD, the top at z=10.83mm is the Beryllium window, and the bottom at z=0.83mm is the Cu-coated GEM.}
\label{fig:z}       
\end{figure}

Tail tracks also have significantly reduced polarization signal compared to peak tracks. The emitted photoelectron is scattered more strongly in the dense GEM and window materials than in the detector gas; as a result, the initial photoelectron direction is often irrecoverable. Fig.~\ref{fig:z} gives the simulated polarization sensitivity as a function of vertical direction along IXPE's GPD, using the moment analysis to estimate photoelectron emission angles. There is a clear drop in sensitivity at both vertical edges of the detector where tail tracks dominate the absorption. Additionally, this figure serves to illustrate the decrease in polarization sensitivity with increasing vertical distance from the pixel array. This is due to charge diffusion in the detector gas blurring tracks before they reach the pixel read-out array (\S\ref{sec:emit}).

If included, tail tracks can cause problems during polarization estimation. Not only do tail tracks degrade the detector polarization sensitivity, the energy-dependent tail fraction can make it difficult to properly calibrate the polarization sensitivity and degrade the energy resolution for continuous spectra. In classical imaging polarimetry, tail tracks are cut from the line source calibration by excluding the low energy tail from the recovered energy histogram, fig.~\ref{fig:tails}. For real, continuous sources this is not possible, so tail tracks remain an important problem for classical polarimetry approaches. \S\ref{sec:removetail} will describe how a NN analysis can identify tail tracks based on their morphological differences and decrease their contamination of the polarization signal.

\subsection{Polarization estimation}
\label{sec:polest}
Polarization estimation can begin once all relevant track features have been recovered. The basic problem is to estimate source polarization parameters $(p_0,\theta_0)$ and their uncertainties from a set of reconstructed emission angles $\{\hat{\theta}_i\}_{i=1}^N$.  As described in the introduction, true emission angles $\theta$ exhibit a sinusoidal modulation with period $\pi$ that depends on the source polarization
\begin{equation} \label{eqn:likelihood}
    p(\theta|p_0,\theta_0) = \frac{1}{2\pi}\big(1 + p_0{\rm cos}\big[2(\theta - \theta_0)\big]\big),
\end{equation}
where $0 \leq p_0 \leq 1$, $-\pi/2 \leq \theta_0 < \pi/2$ and $-\pi \leq \theta < \pi$. Reconstructed emission angles $\hat{\theta}$ are imperfectly recovered, so these follow an adjusted distribution
\begin{equation} \label{eqn:likelihoodmu}
    p(\hat{\theta}|p_0,\theta_0) = \frac{1}{2\pi}\big(1 + p_0\mu{\rm cos}\big[2(\hat{\theta} - \theta_0)\big]\big).
\end{equation}
The modulation factor $0 \leq \mu \leq 1$ is a measure of the detector polarization sensitivity: how well the emission angles are recovered. A higher modulation factor means better polarization sensitivity and $\mu = 1$ means perfect emission angle reconstruction, $\theta = \hat{\theta}$. The modulation factor is the recovered polarization fraction $\hat{p}_0$ for a 100\% polarized source, $p_0 = 1$. Fig.~\ref{fig:mu} gives an example for a $p_0 = 1$ source, the moment analysis recovered emission angles follow the blue $\mu < 1$ distribution while the true emission angles have $\mu = 1$. The modulation factor is a function of photon energy, higher energy tracks have higher modulation factor because they have longer tracks, thus better emission angle estimates. Track reconstruction algorithms and detector hardware both affect the instrument modulation factor. The modulation factor can be considered an average of the reconstruction quality of all tracks for a particular source energy spectrum.  In \S\ref{sec:5} we will revisit the modulation factor from first principles; for now it can be treated as a known constant for a given source energy spectrum, determined by instrument calibration.

\subsubsection{Stokes parameters}
It is usually simpler to estimate the normalized linear Stokes parameters $(\mathcal{Q}, \mathcal{U})$ instead of the polarization fraction and EVPA $(p_0, \theta_0)$. These are an alternative representation of the source linear polarization. Disregarding circular polarization, the Stokes parameters are defined as
\begin{equation}
    Q = Ip_0\cos2\theta_0,
    \label{eqn:q1}
\end{equation}
\begin{equation}
    U = Ip_0\sin2\theta_0,
    \label{eqn:u1}
\end{equation}
where $I$ is the source intensity, and the normalized Stokes parameters are
\begin{equation}
    \mathcal{Q} = p_0\cos2\theta_0,
    \label{eqn:qq1}
\end{equation}
\begin{equation}
    \mathcal{U} = p_0\sin2\theta_0,
    \label{eqn:uu1}
\end{equation}
$-1 \leq \mathcal{Q} \leq 1$, $-1 \leq \mathcal{U} \leq 1$, from which the polarization fraction and EVPA can be derived:
\begin{equation}
    p_0 = \sqrt{\mathcal{Q}^2 + \mathcal{U}^2},
    \label{eqn:p}
\end{equation}
\begin{equation}
    \theta_0 = \frac{1}{2}\arctan\frac{\mathcal{U}}{\mathcal{Q}}.
    \label{eqn:th}
\end{equation}
Using Stokes parameters, the probability density eq.\ref{eqn:likelihoodmu} can be rewritten
\begin{equation}
    p(\hat{\theta}|\mathcal{Q},\mathcal{U}) =  \frac{1}{2\pi} \big(1 + \mathcal{Q}\mu\cos2\hat{\theta} + \mathcal{U}\mu\sin2\hat{\theta} \big).
    \label{eqn:prob_stokes}
\end{equation}
This is a more convenient form because the probability density is linear in the Stokes parameters, unlike for $(p_0, \theta_0)$. Some researchers even prefer to work solely with Stokes parameters. Since likelihoods are invariant to reparameterization, estimating the Stokes parameters and converting back to $(p_0, \theta_0)$ is equivalent to estimating $(p_0, \theta_0)$ directly, although some care must be taken in propagating the uncertainties.

\subsubsection{Methods}
With the modulation factor known, $(p_0, \theta_0)$ and/or $(\mathcal{Q}, \mathcal{U})$ can be recovered from the emission angles $\{\hat{\theta}_i\}_{i=1}^N$ in a number of ways. The simplest method is a binned curve fit, i.e. directly fitting the probability density eq.\ref{eqn:prob_stokes} to the histogram of emission angles $\{\hat{\theta}_i\}_{i=1}^N$, fig.\ref{fig:mu}. Since the Stokes parameters are linear in the probability density eq.\ref{eqn:prob_stokes} they can be fit using binned least squares. The resulting estimates $(\hat{\mathcal{Q}},\hat{\mathcal{U}})$ are unbiased and equivalent to the maximum likelihood estimator (MLE), so long as the bin widths are small enough and each bin contains enough events such that Poisson statistics can be well approximated by Gaussian. This is not always the case, so directly applying the MLE is often a better approach. 

The MLE is the estimator $(\hat{\mathcal{Q}},\hat{\mathcal{U}})$ that maximizes the likelihood function:
\begin{equation}
    L(\{\hat{\theta}_i\}_{i=1}^N|\mathcal{Q},\mathcal{U}) =  \prod_{i=1}^N\frac{1}{2\pi} \big(1 + \mathcal{Q}\mu\cos2\hat{\theta}_i + \mathcal{U}\mu\sin2\hat{\theta}_i \big).
    \label{eqn:likelihood_stoks}
\end{equation}
Here $L$ is a function of $(\mathcal{Q},\mathcal{U})$ and the observations $\{\hat{\theta}_i\}_{i=1}^N$ are fixed. Computationally, it is easier to minimize the negative log-likelihood function
\begin{equation}
    -\log L =  N\log 2\pi - \sum_{i=1}^N \log \big(1 + \mathcal{Q}\mu\cos2\hat{\theta}_i + \mathcal{U}\mu\sin2\hat{\theta}_i \big).
    \label{eqn:loglikelihood_stoks}
\end{equation}
This expression must be minimized numerically to find $(\hat{\mathcal{Q}},\hat{\mathcal{U}})$. Confidence intervals are calculated by numerically integrating the negative log likelihood function in the vicinity of the final estimator. Stokes parameter estimators and their errors can be simply transformed into $(\hat{p}_0,\hat{\theta}_0)$ if desired.

It is often inconvenient to evaluate the MLE and its errors numerically. An analytical solution for the MLE exists if $|\mathcal{Q}\mu| << 1, |\mathcal{U}\mu| << 1$. In practical X-ray polarimetry this limit is nearly always satisfied; polarimeters typically don't achieve much better than $\mu \lesssim 0.5$ averaged over a continuous spectrum and real astrophysical sources generally have low polarization $p_0 \lesssim 0.3$. By applying the Taylor expansion $\log(1+x) \approx x - x^2/2$  to the negative log-likelihood eq.~\ref{eqn:loglikelihood_stoks} and minimizing the resulting quadratic form one finds

\begin{equation}
    \hat{\mathcal{Q}} = \frac{1}{\mu}\frac{(\sum^N_{i=1}\cos2\hat{\theta}_i -  \sum^N_{i=1}\cos2\hat{\theta}_i\sin2\hat{\theta}_i) \sum^N_{i=1}\sin^22\hat{\theta}_i}
    {\sum^N_{i=1}\cos^22\hat{\theta}_i \sum^N_{i=1}\sin^22\hat{\theta}_i - (\sum^N_{i=1}\cos2\hat{\theta}_i\sin2\hat{\theta}_i)^2},
    \label{eqn:qquadform}
\end{equation}

\begin{equation}
    \hat{\mathcal{U}} =  \frac{1}{\mu}\frac{(\sum^N_{i=1}\sin2\hat{\theta}_i -  \sum^N_{i=1}\cos2\hat{\theta}_i\sin2\hat{\theta}_i) \sum^N_{i=1}\cos^22\hat{\theta}_i}
    {\sum^N_{i=1}\cos^22\hat{\theta}_i \sum^N_{i=1}\sin^22\hat{\theta}_i - (\sum^N_{i=1}\cos2\hat{\theta}_i\sin2\hat{\theta}_i)^2}.
    \label{eqn:uquadform}
\end{equation}
Notice that $\mathbb{E}[\sum_{i=1}^N\cos2\hat{\theta}_i\sin2\hat{\theta}_i] = 0$, $\mathbb{E}[\sum_{i=1}^N\cos^22\hat{\theta}_i] = N/2$ and $\mathbb{E}[\sum_{i=1}^N\sin^22\hat{\theta}_i] = N/2$. For the relatively large $N$ in X-ray polarimetry, these terms are effectively constants, so the expressions reduce to
\begin{equation}
    \hat{\mathcal{Q}} = \frac{2}{N\mu} \sum^N_{i=1}\cos2\hat{\theta}_i,
    \label{eqn:qest}
\end{equation}
\begin{equation}
    \hat{\mathcal{U}} = \frac{2}{N\mu} \sum^N_{i=1}\sin2\hat{\theta}_i.
    \label{eqn:uest}
\end{equation}
Because these are unbiased and MLE, they are the minimum variance unbiased estimators for $(\mathcal{Q}, \mathcal{U})$. In other words, so long as $|\mathcal{Q}\mu| \lesssim < 1, |\mathcal{U}\mu| \lesssim < 1$ and $N \gtrsim 1000$, these are the best possible estimators for $(\mathcal{Q}, \mathcal{U})$.

\begin{figure}[t]
\centering
\includegraphics[scale=.45]{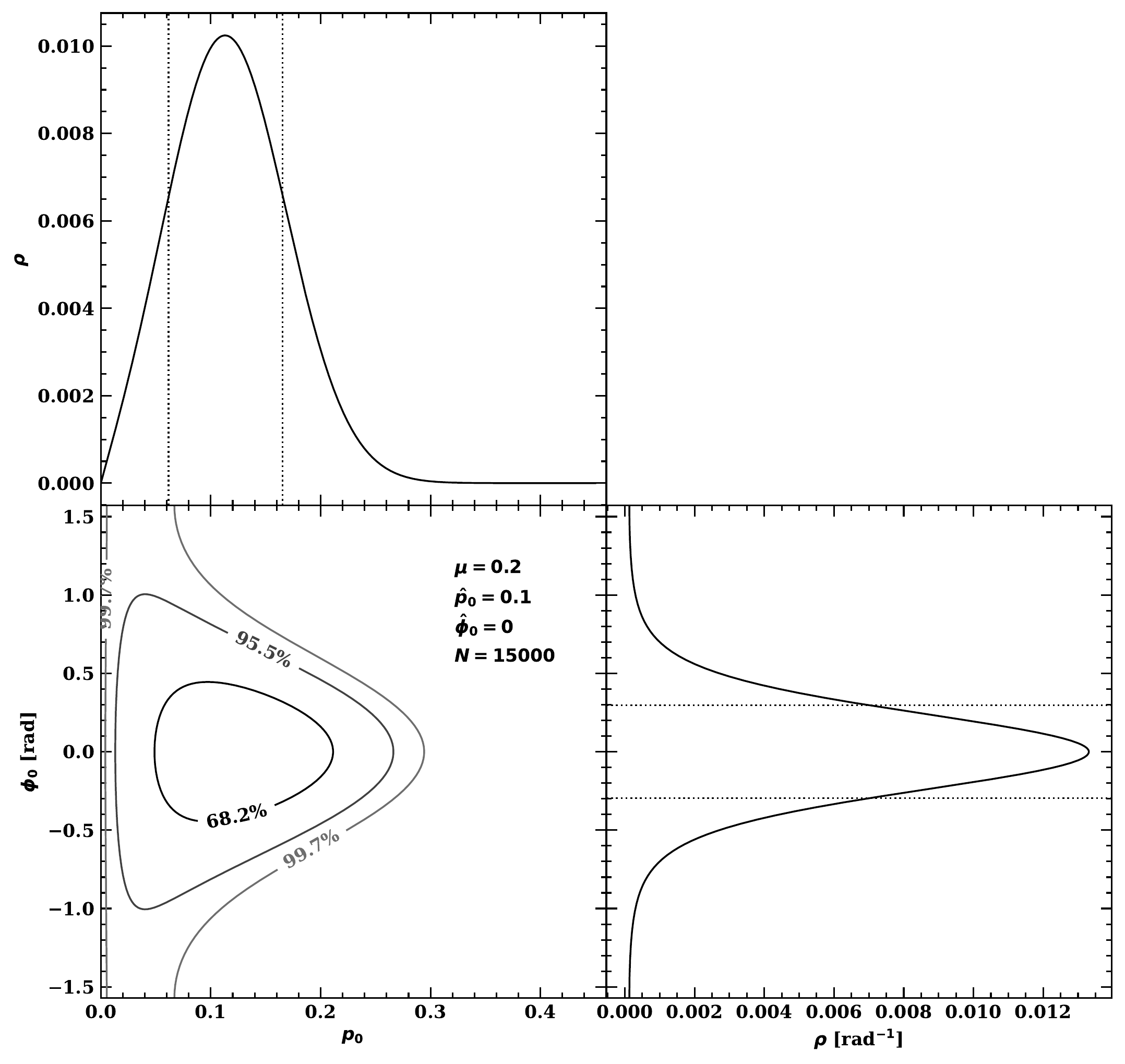}
\caption{Confidence intervals from full posterior distribution (eq.\ref{eqn:posterior}) over polarization parameters $(p_0, \phi_0)$ given estimates $(\hat{p}_0, \hat{\phi}_0)$ using $N$ events and instrument modulation factor $\mu$. 
Confidence intervals for $68.2\%$,$95.5\%$ and $99.7\%$ (1$\sigma$, $2\sigma$, $3\sigma$) are displayed.
The full marginal posterior distributions are plotted on the wings with $68.2\%$ (1$\sigma$) confidence intervals drawn. }
\label{fig:post}       
\end{figure}

Kislat et al. \cite{kislat_analyzing_2015} derive the full analytical posterior distribution for both $(\mathcal{Q},\mathcal{U})$ and $(p_0,\theta_0)$ given the estimators eqs.\ref{eqn:qest},\ref{eqn:uest}. They find the $(\mathcal{Q},\mathcal{U})$ posterior follows a bivariate normal distribution while the $(p_0,\theta_0)$ posterior is
\begin{equation}
\begin{aligned}
    p(p_0,\theta_0 | \hat{p}_0,& \hat{\theta}_0) ={} \frac{\sqrt{N}\hat{p}_0 \mu^2}{2\pi\sigma} \times \\
    &\exp\Bigg[-\frac{\mu^2}{4\sigma^2}\Bigg\{ \hat{p}_0^2 + p_0^2 - 2\hat{p}_0 p_0\cos(2(\hat{\theta}_0 - \theta_0)) \\ & - \frac{\hat{p}_0^2p_0^2\mu^2}{2}\sin^2(2(\hat{\theta}_0 - \theta_0)) \Bigg\} \Bigg],
\end{aligned}
\label{eqn:posterior}
\end{equation}
where 
\begin{equation}
    \sigma = \sqrt{\frac{1}{N} \left( 1 - \frac{p_0^2\mu^2}{2}\right)}.
    \label{eqn:sig}
\end{equation}
The posterior assumes a uniform prior over $(p_0,\theta_0)$. With the posterior in hand, any desired confidence interval can be computed analytically or numerically. 
High $\mu$ and high $N$ reduce the width of the posterior so are both desirable to minimize the errors on recovered polarization parameters.
Note that the estimator $\hat{p}_0$ is not unbiased, its posterior is highly asymmetric for low $p_0$ since $p_0 \geq 0$. For $p_0 = 0$ some polarization fraction $\hat{p}_0 > 0$ will always be measured. The amount of polarization likely to be measured when $p_0 = 0$ depends on the width of the posterior, i.e. $N$ and $\mu$, and is often used as a measure of detector sensitivity or signal-to-noise ratio (SNR), \S\ref{sec:mdp}.
Fig.~\ref{fig:post} shows the posterior distribution for an example observation.

The recommended approach to polarization estimation is using eqs.~\ref{eqn:qest}, \ref{eqn:uest} for point estimates and the derived posterior eq.~\ref{eqn:posterior} for confidence intervals. This approach is expedient and optimal given the instrumental constraints of imaging X-ray polarimetry. In \S\ref{sec:5} we extend eqs.~\ref{eqn:qest} -- \ref{eqn:posterior} to include neural network uncertainty estimates on predicted emission angles $\hat{\theta}$.


\subsubsection{Minimum detectable polarization (MDP)}
\label{sec:mdp}

The standard figure-of-merit used in X-ray polarimetry to compare instrument sensitivity is minimum detectable polarization (MDP) \cite{weisskopf_understanding_2010}. MDP$_{99}$ is the polarization fraction that has a 1\% probability of being exceeded by chance for an unpolarized ($p_0 = 0$) source. This can be found by integrating the posterior distribution eq.\ref{eqn:posterior}:
\begin{equation}
    \int_0^{\rm MDP_{99}} \int_{-\pi/2}^{\pi/2} p(p_0, \theta_0 | \hat{p}_0, \hat{\theta_0})d\theta_0 dp_0 = 0.99,
\end{equation}
\begin{equation}
    {\rm MDP}_{99} = \frac{4.29}{\mu\sqrt{N}}.
\end{equation}
On fig.~\ref{fig:post}, the MDP$_{99}$ would be the one sided $99\%$ confidence interval for the $p_0$ marginal distribution (top panel).

The MDP$_{99}$ depends only on the modulation factor (signal) and the Poisson counting noise $1/\sqrt{N}$; it is effectively an inverse of the SNR. Polarimeters with lower MDP$_{99}$ can expose for shorter times and get the same confidence on their polarization measurements. Track reconstruction and polarization estimation approaches that increase the modulation factor for a fixed number of tracks $N$ will decrease the MDP$_{99}$. Sections \ref{sec:4} and \ref{sec:5} will describe how a neural network approach can improve both track reconstruction and polarization estimation to minimize the MDP$_{99}$.

\newpage
\section{Deep neural networks}
\label{sec:3}
Deep neural networks have achieved state-of-the-art performance on a wide variety of machine learning tasks and are becoming increasingly popular in domains such as speech recognition \cite{graves_connectionist_2006}, natural language processing \cite{young_recent_2018}, bioinformatics \cite{tang_recent_2019} and especially computer vision \cite{krizhevsky_imagenet_2012}. Going from track images to emission angle estimates can be classified as a computer vision problem, so it is not surprising that neural networks would be well suited to track reconstruction. The Cherenkov Telescope Array (CTA) \cite{brill_investigating_2019} team have applied related deep learning methods to differentiate between cosmic rays and gamma rays. Notably they also have to deal with a hexagonal pixel grid, also found in imaging X-ray polarimeters. The IceCube collaboration has begun the use of graph neural networks to identify 3D neutrino tracks with great success \cite{choma_graph_2018}. 
This section briefly covers deep neural network and machine learning concepts relevant to X-ray polarimetry, but for a proper introduction to this important field see \cite{goodfellow_deep_2016, hastie_elements_2009}.

\subsection{Machine learning with deep neural networks}
\label{sec:ML}
Deep neural networks (DNNs) are a type of supervised machine learning algorithm. Machine learning algorithms learn from data, as opposed to a fixed algorithm programmed by human beings. Supervised machine learning algorithms learn from labelled data: known input-output pairs $\{x_i,y_i\}^N_{i=1}$. An example from X-ray polarimetry would be track image (input), emission angle (output). Supervised machine learning algorithms can be used to solve a variety of tasks. Most tasks belong to one of two groups: regression or classification. In regression the algorithm is trained to predict a continuous numerical value given an input. In classification, the algorithm predicts the probability an input belongs to a particular class from a preselected group of finite classes. In the case of track reconstruction, we are interested mainly in regression.

In essence, DNNs approximate the unknown function $f^{\star}$ that maps problem inputs to outputs $y = f^{\star}(x)$. DNNs are organized as a forward network of function layers each with their own learnable parameters or weights $w$, e.g. $f(x) = f_3(f_2(f_1(x;w_1);w_2);w_3)$. The initial function layer $f_1$ is known as the input layer, the final layer $f_3$ is the output layer, and all layers in between are the hidden layers. Using many hidden layers tends to improve the predictive capacity of the DNN, whence the term 'deep learning'. During training, the DNN $f(x)$ is iteratively directed towards $f^{\star}(x)$.

\begin{figure}[t]
\centering
\includegraphics[scale=.35]{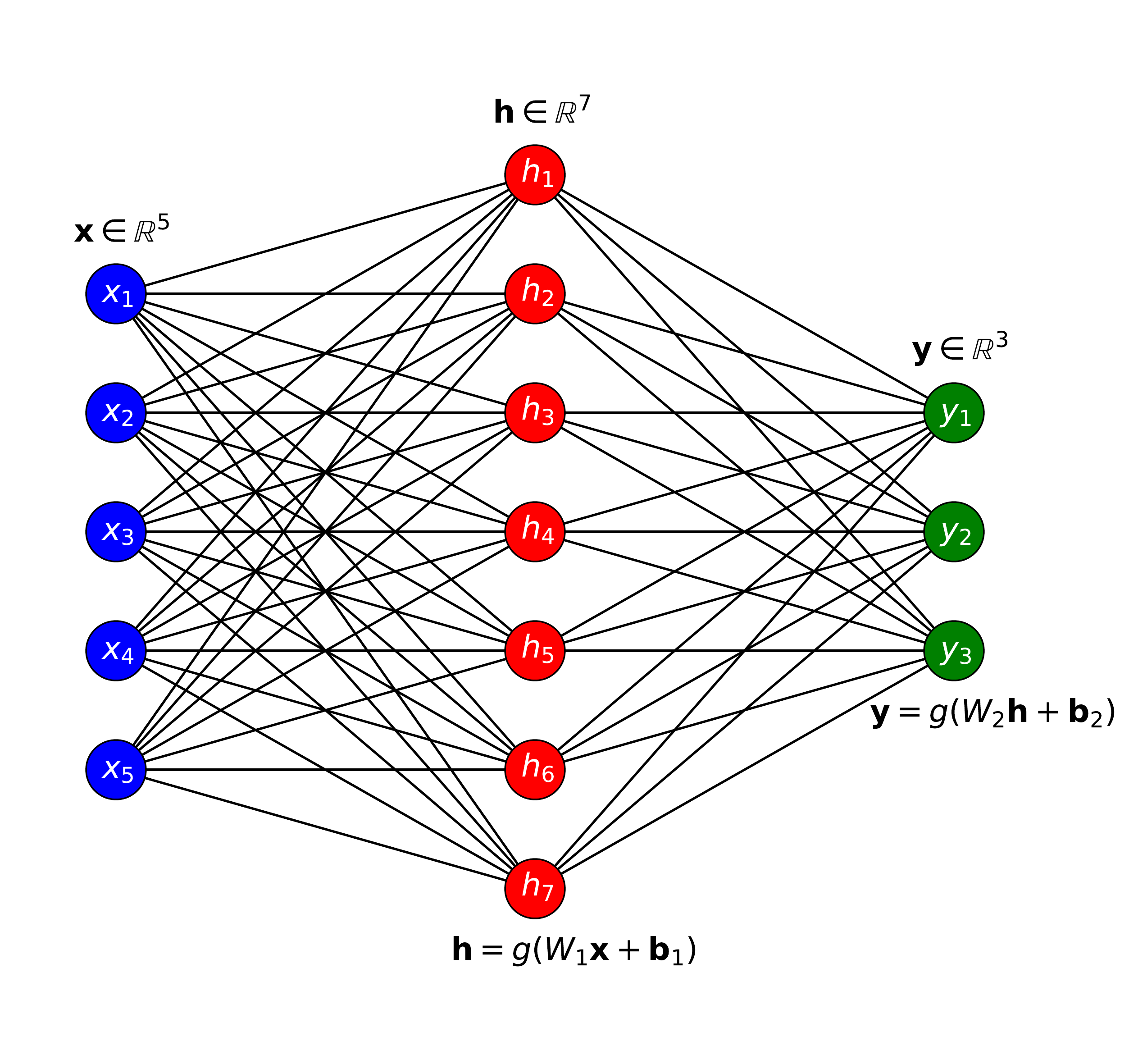}
\caption{Feed-forward neural network with one hidden layer. The neural network takes vector $\mathbf{x}$ as input and produces vector $y$ as output.}
\label{fig:net}       
\end{figure}

If the layer functions $f_1, f_2, f_3$ were linear then the DNN $y = f(x)$ would be a linear model $y = w^Tx$, equivalent to a simple least squares approach. DNNs use non-linear layer functions. This allows the DNN to learn non-linear relationships between the input features $y = w_2^Tf_1(x; w_1)$. Here $f_1(x;w_1)$ is a non-linear function with learnable parameters $w_1$.
In this way DNNs are able to learn what functions of the input features should be combined, unlike a linear model that linearly combines input features directly.
A simple example of a non-linear layer function used in DNNs is $f_1(x;w_1,b) = g(w_1^Tx + b)$ where $w_1$ is a matrix and $g$ is a non-linear function such as $\tanh$, the sigmoid function or a rectified linear unit (ReLU). DNNs using this kind of layer are universal function approximators in the asymptotic limit \cite{hornik_multilayer_1989}, so they should in principle well approximate any $f^{\star}(x)$. In practice, memory, training and data limitations make this difficult to achieve. Many DNNs use domain specific layer functions and architectures to reduce these limitations and converge to $f^{\star}(x)$ more quickly, see \S\ref{sec:CNN}. Figure \ref{fig:net} shows a schematic of a simple DNN.

\subsubsection{Training}
\label{sec:training}
As in supervised linear models like least squares regression, DNNs are trained to minimize a specific loss function over the observed data. In most tasks this normally means minimizing the negative log-likelihood, as in \S2.2. For regression tasks, if the output distribution family $p(y|x)$ is unknown, a Gaussian should usually be the default since it is the least informative choice. For a Gaussian likelihood, the negative log likelihood is 
\begin{equation}
    L(\{y_i,x_i\}_{i=1}^N\mid\mathbf{w}) \propto \sum^N_{i=1}\|y_i - \hat{y}(x_i|\mathbf{w})\|^2,
    \label{eqn:egloss}
\end{equation}
the mean squared error (MSE) of prediction for the dataset. The loss function $L(\mathbf{w})$ is minimized over the DNN weights $\mathbf{w}$. MSE is a common loss function for regression with DNNs.

For classification tasks, $p(y|x)$ is usually assumed to follow a Bernoulli (in binary, 2-class classification) or categorical distribution (in >2 class classification). In the Bernoulli case, the negative log-likelihood is
\begin{equation}
    L(\{y_i,x_i\}_{i=1}^N\mid\mathbf{w}) \propto \sum^N_{i=1} - y_i\log\big(\hat{y}(x_i|\mathbf{w})\big) - (1 - y_i)\log\big(1 - \hat{y}(x_i|\mathbf{w}) \big).    
    \label{eqn:eglossbin}
\end{equation}
In order to ensure $0 \leq \hat{y}_i \leq 1$, DNNs for classification usually end with a softmax layer \cite{bridle_training_1990}. These project DNN outputs onto the appropriate simplex, i.e., makes sure all predicted probabilities sum to one. For K output classes
\begin{equation}
    \label{eqn:softmax}
    {\rm Softmax}(\mathbf{x})_i = \frac{e^{x_i}}{\sum_{i=1}^K e^{x_i}}
\end{equation}

Since DNNs use non-linear functions, eq.~\ref{eqn:egloss} or \ref{eqn:eglossbin} cannot be minimized analytically. Non-linearity makes the loss function complex with multiple local minima and saddle points in a very high dimensional weight space $\{\mathbf{w}\}$. 
This challenging minimization problem is usually tackled with variations on the stochastic gradient descent (SGD) algorithm \cite{goodfellow_deep_2016}. At each iteration SGD approximates the gradient of the loss function using a subset of the training data
\begin{equation}
    g_i = \nabla_{\mathbf{w}} L(\mathbf{w}) \approx \sum_{i=1}^n \nabla_{\mathbf{w}}\|y_i - \hat{y}(x_i|\mathbf{w})\|^2, ~~n << N
\end{equation}
and takes a step in parameter ($\mathbf{w}$) space along the direction of the negative gradient
\begin{equation}
    \mathbf{w}_{i+1} = \mathbf{w}_{i} - \epsilon g_i.
\end{equation}
The process is repeated until suitable convergence. The weights are initialized randomly \cite{} and the learning rate $\epsilon$ is a tuneable scalar for the specific application. The gradient for each individual $w_i$ can be calculated by using the chain rule through each layer of the DNN, in a process called backpropagation \cite{goodfellow_deep_2016}. In order for SGD to work, the DNN has to use a differentiable loss function and layer functions $f_1,f_2,f_3$. 
The whole dataset $N$ is not used to calculate the gradient because it is too slow and often too large to fit into memory. Furthermore, DNNs benefit from using small batches $n$ to approximate the gradient because the added stochasticity lets the minimization escape saddle points and local minima \cite{goodfellow_deep_2016}. Usually DNNs are trained on graphical processing units (GPUs), specialized computer hardware for fast and parallelizable linear algebra operations. 

\subsubsection{Validation and model selection}
\label{sec:val}
DNNs typically have on the order of millions, sometimes billions \cite{brown_language_2020}, of trainable weights $\mathbf{w}$. The number of weights is often higher than the number of training examples, making overfitting the labelled dataset likely. This is the central challenge of machine learning: the algorithm must perform well on new inputs, not just those in the training dataset. To ensure model generalization to unseen examples, part of the training data is set aside for validation. 
The trainable weights $\mathbf{w}$ are not trained on these examples.
DNNs come with a number of additional model parameters -- hyperparameters -- that can be adjusted to improve model generalization; examples include the learning rate $\epsilon$, the neural network architecture (type and number of layers) and any regularization. During training, the DNN is evaluated on the validation examples to tune the hyperparameters. This process is known as model selection. 

Regularization is an important hyperparameter for improving model generalization. In exactly the same way as linear models can be regularized, for example in ridge regression, one can regularize a DNN by adding a penalty on the weights $\mathbf{w}$ to the loss function. A common example is L2 regularization:  
\begin{equation}
    L(\mathbf{w}) \propto \lambda\|\mathbf{w}\|^2 + \sum^N_{i=1}\|y_i - \hat{y}(x_i|\mathbf{w})\|^2.
    \label{eqn:regloss}
\end{equation}
The additional term restricts the model space by encouraging the weights $\mathbf{w}$ to be close to zero. The hyperparameter $\lambda$ can be tuned to maximize generalization on the validation examples. This type of regularization is equivalent to a Gaussian prior on the weights with mean zero. By restricting the model space, regularization helps prevent overfitting and improves generalization when $\lambda$ is appropriately tuned.

\subsection{Convolutional Neural Networks}
\label{sec:CNN}
Some DNNs have specialized layer functions that provide an inductive bias helpful for specific inputs. Convolutional neural networks (CNNs) \cite{lecun_convolutional_1998} were designed for processing inputs that have neighbouring position correlations, for example 2D images, where adjacent pixel values are often correlated, or 1D time series where adjacent time points are often correlated. CNNs use layer functions that apply a convolution to the input. The convolution kernel function is learned during training. This kind of layer function is invariant under translation of the input, an important prior for working with image data.
CNNs have revolutionized image processing and computer vision applications since winning the Imagenet benchmark competition in 2012 \cite{krizhevsky_imagenet_2012}.  
(Paragraph and plot of convolution here maybe)

Recently, a certain class of CNNs called residual networks (ResNets) have achieved state-of-the-the-art results in multiple image benchmark tasks \cite{he_deep_2015}. ResNets introduce 'skip' connections between certain layers that allow the CNN to ignore the layers in between. This alleviates many of the difficulties in training extremely deep neural networks.

\subsection{Multitask learning}
\label{sec:multi}
So far, only DNNs with single objectives and outputs have been considered. However, DNNs can be trained to solve multiple tasks at once, producing a vector of outputs. Solving multiple tasks at once often performs better than solving each task individually. This is partly because multiple tasks have a regularizing effect on each other, penalizing model complexity and overfitting on any single task leading to better overall model generalization \cite{zhang_survey_2021}. Moreover, since each task shares the same input representation, e.g., an image, the features learned for a single task can help the others train faster and better.

In a multitask learning setting, the loss function includes the terms for each individual task. For example, in regression with three tasks and an MSE loss for each
\begin{equation}
    L(\mathbf{w}) \propto \sum^N_{i=1}\|y^1_i - \hat{y}^1(x_i|\mathbf{w})\|^2 + \alpha\sum^N_{i=1}\|y^2_i - \hat{y}^2(x_i|\mathbf{w})\|^2 + \beta\sum^N_{i=1}\|y^3_i - \hat{y}^3(x_i|\mathbf{w})\|^2,
    \label{eqn:megloss}
\end{equation}
where the DNN outputs concatenated vector $(\hat{y}^1,\hat{y}^2,\hat{y}^3)$.
The relative size of each task loss term is controlled by the hyperparameters $\alpha,\beta$. As usual, these hyperparameters should be tuned to maximize the model performance on unseen examples. More often in multitask learning, each task has a different form of loss function.

\subsection{Uncertainty quantification}
Once trained, DNNs can achieve state-of-the-the-art prediction accuracy in many domains. However, these models cannot capture the uncertainty inherent in their predictions. Predictive uncertainty quantification can be crucial, especially in applications where the real-world data distribution differs from the training data distribution. In this scenario, DNNs can extrapolate, producing overconfident but highly inaccurate predictions sometimes with disastrous consequences \cite{amodei_concrete_2016}.
Predictive uncertainty is best represented by a posterior distribution on the predicted parameters given the inputs, however, simply quoting the variance of the posterior, often assumed to be Gaussian, is common practice.

There are two germane types of uncertainty one can model \cite{kendall_what_2017}. \textit{Aleatoric uncertainty} captures noise inherent in the data. This is equivalent to statistical uncertainty or unexplained variance. On the other hand, \textit{epistemic uncertainty} accounts for uncertainty in the model parameters – uncertainty which captures our ignorance about which model generated our collected data. This uncertainty can be reduced given the appropriate choice of model or enough data, and is often referred to as model uncertainty, systematic uncertainty or explained variance. More explicitly \cite{choi_uncertainty-aware_2017}, training a model $\hat{y} = f(\mathbf{x})$ to approximate $y = f^{\star}(\mathbf{x}) + \epsilon$, where the statistical noise $\epsilon$ is normally distributed $\epsilon \in \mathcal{N}(0,\sigma^2_a)$, the expected error is
\begin{equation}
\begin{aligned}
    \mathbb{E}\|y - \hat{y}\|^2 &= \mathbb{E}\|y - f^{\star}(\mathbf{x}) + f^{\star}(\mathbf{x}) - f(\mathbf{x})\|^2 \\
    &=  \mathbb{E}\|y - f^{\star}(\mathbf{x})\|^2 + \mathbb{E}\|f^{\star}(\mathbf{x}) - f(\mathbf{x})\|^2 \\
    &= \sigma_a^2 + \sigma_e^2,
    \label{eqn:aleaepi}
\end{aligned}
\end{equation}
where $\sigma_a^2$ and $\sigma_e^2$ are the aleatoric and epistemic errors, respectively. Cross terms in eq.~\ref{eqn:aleaepi} go to zero since $\epsilon$ is independent of $y$ and $\hat{y}$. Eq.~\ref{eqn:aleaepi} highlights how epistemic uncertainties arise from a systematic difference between the true and learned models, while aleatoric uncertainties are irreducible statistical fluctuations about the true model.

Aleatoric uncertainties can be further broken down into two types: homoskedastic and heteroskedastic uncertainties. Homoskedastic uncertainties are the same for all inputs, whereas heteroskedastic uncertainties depend on the input $\mathbf{x}$, so $\epsilon \in \mathcal{N}(0,\sigma^2_a(\mathbf{x}))$. Almost all real data has heteroskedastic uncertainty; some examples are better measured than others. Photoelectron tracks are extremely heteroskedastic. 

Uncertainty quantification in DNNs is a quickly developing field with many possible approaches. Modelling aleatoric uncertainties is simpler and a similar method is shared between approaches. This involves training the DNN to predict its own aleatoric uncertainty by enhancing the loss function. The next section, \S\ref{sec:deepens}, provides a concrete example. Epistemic uncertainty is more difficult to quantify since it usually involves sampling over different models. Bayesian neural networks \cite{kendall_what_2017} do this by randomly subsampling the trained DNN weights $\mathbf{w}$ and making multiple predictions, while deep Gaussian processes \cite{jakkala_deep_2021} or mixture density networks \cite{choi_uncertainty-aware_2017} model the epistemic uncertainty intrinsically. For practical applications, which require robust and scaleable uncertainty predictions as well as high prediction accuracy, deep ensembles \cite{lakshminarayanan_simple_2017} are the state of the art \cite{gustafsson_evaluating_2020}.

\subsubsection{Deep ensembles}
\label{sec:deepens}

\begin{figure}[t]
\centering
\includegraphics[scale=.28]{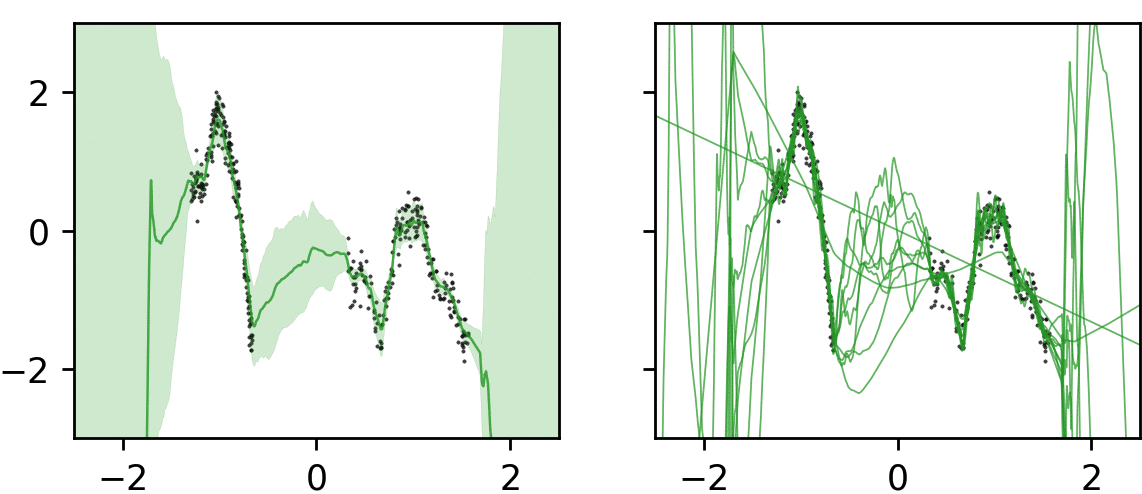}
\caption{Toy regression problem using a deep ensemble. Right panel shows individual DNN prediction functions. Left panel gives the estimated epistemic uncertainty of the deep ensemble. Figure from \cite{antoran_depth_2020}.}
\label{fig:epinet}       
\end{figure}

Deep ensembles are made up of an ensemble of individual DNNs, each trained independently on the same dataset to predict the desired output features. Different random initializations of the same NN at the start of training leads to widely different prediction functions \cite{fort_deep_2019}. Deep ensembles exploit this property by incorporating the results of many differently initialized NNs, increasing the diversity of predictors. This not only yields higher prediction accuracy, but also allows for state-of-the-art aleatoric and epistemic uncertainty quantification \cite{ovadia_can_2019, gustafsson_evaluating_2020}.

To model the aleatoric uncertainty in a regression task, each individual DNN composing the ensemble is trained to minimize the following negative log-likelihood loss function over all training examples $i$ \cite{lakshminarayanan_simple_2017}
\begin{equation}
    L(y_i\mid\mathbf{x}_i) = \frac{{\rm log}(\hat{\sigma}^2(\mathbf{x}_i))}{2} + \frac{\|y_i - \hat{y}(\mathbf{x}_i)\|_2^2}{2\hat{\sigma}^2(\mathbf{x}_i)}.
\end{equation}
The DNN predicts the aleatoric uncertainty $\hat{\sigma}_a$ for each example prediction, as well as the example label $\hat{y}$ as in standard DNN regression, eq.~\ref{eqn:egloss}. This form of loss function assumes the aleatoric error has a Gaussian distribution, usually a reasonable approximation. In practice the DNNs are trained to predict the log variance $\hat{s}(\mathbf{x}_i) = {\rm log}(\hat{\sigma}^2(\mathbf{x}_i))$
\begin{equation}
\label{eqn:log_loss}
    L(y_i\mid\mathbf{x}_i) = \frac{\hat{s}(\mathbf{x}_i)}{2} + \frac{e^{-\hat{s}(\mathbf{x}_i)}\|y_i - \hat{y}(\mathbf{x}_i)\|_2^2}{2}
\end{equation}
since this is more numerically stable \cite{kendall_what_2017}.

During model prediction, predictions from all DNNs composing the ensemble are combined. The final predicted label is the mean over the ensemble predictions. For an ensemble with $M$ DNNs $\hat{y}(\mathbf{x}) = (1/M)\sum_{j=1}^M \hat{y}^j(\mathbf{x})$. Similarly, the final aleatoric uncertainty is given by $\hat{\sigma}^2_a(\mathbf{x}) = (1/M)\sum_{j=1}^M \hat{\sigma}_a^2(\mathbf{x})^j$.
The variance of the label predictions $\hat{y}^j$ over the ensemble is used to model the epistemic uncertainty, resulting in total predicted uncertainty
\begin{equation}
\begin{aligned}
    \hat{\sigma}^2(\mathbf{x}) &= \hat{\sigma}^2_a + \hat{\sigma}^2_e \\
    \hat{\sigma}^2(\mathbf{x}) &= \frac{1}{M}\sum_{j=1}^M \hat{\sigma}_a^2(\mathbf{x})^j + \frac{1}{M}\sum_{j=1}^M\Big[\hat{y}^j(\mathbf{x})  - \frac{1}{M}\sum_{j=1}^M \hat{y}^j(\mathbf{x})\Big]^2.
\end{aligned}
\end{equation}
Very few DNNs are needed in the ensemble, $M \approx 5 - 10$, for the method to achieve state-of-the-art uncertainty quantification \cite{lakshminarayanan_simple_2017}. Figure \ref{fig:epinet} visualizes the epistemic uncertainty prediction for a toy regression problem. In the area far from the training examples, the DNNs composing the ensemble strongly disagree in their predictions, giving a high epistemic uncertainty.

In multitask problems, the aleatoric uncertainties can actually replace the hyperparameters that determine the relative importance of individual tasks in the loss function \cite{kendall_multi-task_2018}. The DNNs automatically tune their own hyperparameters by learning the aleatoric uncertainties during training. 

\newpage
\section{Neural networks for track reconstruction}
\label{sec:4}

Track reconstruction is an image input, multi-output regression problem \S\ref{sec:trckrec}. From a machine learning perspective, this problem is perfectly suited to a multitask CNN approach. CNNs learn translation invariant representations \S\ref{sec:CNN}, an important constraint for working with image inputs, and a multitask approach is not only more computationally efficient but also likely to improve the individual task performances \S\ref{sec:multi}. In practice, a deep ensemble of CNNs is better than a single CNN for track reconstruction:
\begin{itemize}
    \item An ensemble of models provides better predictive performance and generalization, \S\ref{sec:deepens}.
    \item Trustworthy uncertainty quantification will be important for the next step: polarization estimation, \S\ref{sec:5}.
    \item Deep ensembles provide automatic hyperparameter tuning of individual task importance for the multitask loss function, \S\ref{sec:deepens}.
\end{itemize}

Some authors \cite{kitaguchi_convolutional_2019} have argued for an end-to-end deep learning approach to polarization estimation that combines track reconstruction with polarization estimation into one huge supervised learning problem to go directly from a set of tracks to source polarization. Although initially attractive because one could use source polarization directly as the neural network learning signal, avoiding the need for simulated detector events, this approach is difficult. Training would be costly (perhaps infeasible) since many individual events would be required for a single training example. Moreover, the observer may wish to adjust partitioning of the data set into e.g., time, energy, and spatial subsets, with differing polarization. The best partition will often not be obvious before the analysis starts, thus combining the properties of individual tracks allows a more efficient exploration of binning options. Finally detector-dependent artifacts can best be handled from individual tracks (with fine spatial positioning), rather than point spread function (PSF) weighted groups of tracks.

The neural network approach presented in this section and the next splits imaging X-ray polarimetry into two mostly separate data analysis steps, as described in \S\ref{sec:2}. This section describes step 1, track reconstruction \S\ref{sec:trckrec}: extracting the relevant features (emission angles, absorption points, energies, and their associated uncertainties) from photoelectron track images as well as possible. This section follows the current state-of-the-art deep ensemble approach for track reconstruction taken by \cite{peirson_deep_2021, peirson_towards_2021}.

\subsection{Dataset}
\label{sec:data}
In any supervised learning problem, a dataset of labelled examples is an essential first step, without this, training a model is not possible. In the case of track reconstruction, one needs photoelectron track images labelled with known photoelectron 2D emission directions, 2D absorption points on the detector grid and scalar photon energies. Unfortunately, these labelled events cannot be collected from the detector in the lab so must be simulated. Most telescope detectors have event simulators to study detector performance; for example, IXPE has a GPD Monte Carlo event simulator built in GEANT4 \cite{agostinelli_geant4simulation_2003} as part of its software suite. In any application where simulated data is used to train a machine learning model, care must be taken to make sure the simulated training data truly matches the real data the algorithm is likely to see. Systematic difference between the two is known as covariate shift and can cause reductions in prediction accuracy and uncertainty quantification on real data compared to the simulated data. In the case of IXPE, simulated events have been rigorously verified to match with real lab-collected detector events \cite{baldini_design_2021} at the expected telescope settings, including photons absorbed outside the gas volume \S\ref{sec:tail}. For IXPE trained neural network approaches, the performance on polarization estimation for simulated events generalizes to real detector data \cite{peirson_deep_2021, peirson_towards_2021}, suggesting minimal covariate shift.

Once labelled photoelectron track events can be reliably simulated for the X-ray polarimeter of choice, one needs to choose the amount and type of training and testing data to be simulated. For IXPE's GPDs, \cite{peirson_deep_2021, peirson_towards_2021} use a training set of approximately 3.3 million tracks simulated uniformly for an unpolarized source between 1 - 11keV, for performance across a broad spectrum. They keep separate an additional 200k simulated tracks for validation and testing. The energy distribution of the training set does not make much difference to track reconstruction or polarization prediction, so long as no energies are greatly underrepresented nor labeled energy bins significantly larger than the energy resolution. An unpolarized source is simulated to make sure emission angles are uniformly distributed, so as not to cause any learning bias. The next section describes additional data pre-processing necessary for non-square grids.
\begin{svgraybox}
In general, the larger the DNN (in terms of parameters and layers), the more data is required for training.
\end{svgraybox}

\begin{figure}
\centering
\includegraphics[width=1.05\textwidth]{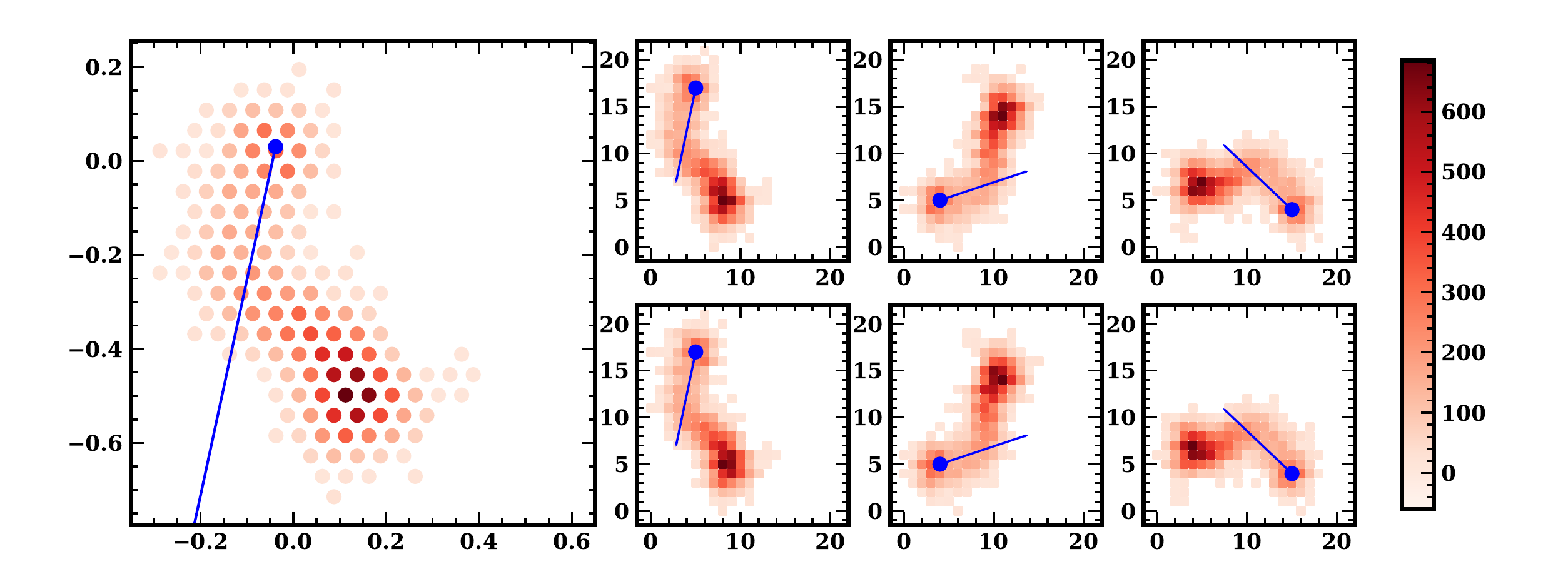}
\caption{Example square conversions of a 6.4 keV hexagonal track (left panel). The six panels to the right show shifts along the 120$^\circ$ GPD axes; shifting odd rows (upper) or even rows (lower). For each hexagonal track, CNNs are fed column-wise pairs of square conversions, along with the energy, absorption point (blue dot) and initial photoelectron direction (blue line) as labels.}
\label{fig:square}
\end{figure}

\subsection{Geometric bias}
\label{sec:geom}
Imaging polarimeters like IXPE use hexagonal pixels for two reasons: 
\begin{enumerate}
    \item Hexagonal arrays are the densest possible 2D packing, meaning more pixels per unit area thus improved track resolution.
    \item Hexapolar grid effects are orthogonal to the quadrupolar polarization signal. 
\end{enumerate}
The discrete nature of the pixel grid means that track elongations and emission directions are better resolved along some axes of symmetry than others, potentially creating preferred directions for reconstruction. This bias would only be visible when the pixel size is significant compared to the track size, i.e., for small, low energy tracks.
The hexagonal grid used in the IXPE GPDs is designed to minimize any polarization systematics arising from these biases.
Although not all imaging polarimeters use hexagonal pixels, this section describes some potential neural network solutions for dealing with non-square pixels.

A hexagonal grid is not natively compatible with standard CNN implementations for square matrices. It is possible to transform from a hexagonal to a square grid, however a naive transformation can lead to polarization systematics. Polarization estimation, step 2 \S\ref{sec:6}, expects unbiased estimators for the emission angles. CNNs learn to prefer some emission angles over others, possibly interfering with the polarization signal.
In a naive hexagonal to square conversion, emission angle bias happens partly because the CNN convolutional kernels are not spatially equivariant in hexagonal space. A possible solution is to work directly with hexagonal convolutions \cite{steppa_hexagdly_2019}, but \cite{peirson_deep_2021} describe an expedient method.


\subsubsection{Hexagonal to square conversion}
It is possible to leverage existing square CNN architectures by using an appropriate polarization-systematic-free hexagonal to square conversion. From \cite{peirson_deep_2021}:
\begin{quotation}
There are two main ways of converting between hexagonal and square grids: interpolation and pixel shifting. Interpolation places a fine square grid on top of the hexagonal image and interpolates.
Interpolation should generally be avoided since it adds noise to the raw data and is not easily reversible.
Pixel shifting rearranges pixels by shifting alternate rows and then rescaling \cite{steppa_hexagdly_2019} ... To avoid any polarization bias from the conversion, we pixel shift each track along each of the six hexagonal axes. Hexagonal tracks are rotated so that rows align horizontally (this can be done in three different ways, separated by $120^{\circ}$), then alternate rows are shifted (this can be done in two different ways, left and right) so that the track resembles a rectangular grid. We convert the rectangular grid into a square image by defining the leftmost track pixel and bottom track pixel as the left edge and the base of the image respectively. A square image size of 50x50 pixels is used to fit all track sizes for energies up to 11 keV.
Since square track images are defined independently of the absolute hexagonal coordinate values, the initial hexagonal track rotation can be performed about any axis.


A single hexagonal track thus produces six square conversions, fig.~\ref{fig:square}, two for each $120^{\circ}$ angle.
A single training example for the CNNs is formed by stacking the corresponding square conversion pair, similarly to color channels in a \textit{rgb} image CNN problem -- in this case with only two channels. At test time all 3 pairs are evaluated by the CNNs and the predicted angles are rotated back to their original direction. 
\end{quotation}
This approach artificially approximates spatial equivariance of CNN convolution kernels in the hexagonal space and removes all relevant prediction bias on emission angles $\theta$ (and later $p_0, \theta_0$) introduced when converting from hexagonal to square coordinates.



\subsection{Deep ensemble setup} 
\label{sec:deepset}
The specific CNN model that will compose the deep ensemble has to be chosen before any training begins. CNN architectures are usually compared by their performance on natural image classification datasets, like imagenet (224x224 pixels) or CIFAR (32x32 pixels) \cite{doon_cifar-10_2018}. The residual network architecture (ResNet \cite{he_deep_2015}) described in \S\ref{sec:CNN} and related architectures like DenseNet \cite{huang_densely_2018} became the state of the art for image processing benchmarks in 2017. \cite{peirson_deep_2021} use the ResNet-18 CNN architecture to compose the deep ensemble; this has 18 total layers. The results shown in this section are using ResNet-18s, but any DNN appropriate for images could be used. Indeed, as of 2020, transformer architectures \cite{vaswani_attention_2017} have supplanted residual networks as the latest state of the art in image processing. 

During training, each CNN minimizes a loss function. Since track reconstruction is a multitask problem, the loss function will include multiple terms for emission angle loss, absorption point loss, and energy loss. The DNNs will also predict the aleatoric uncertainty on each of its estimates. Each task loss function term is considered separately and they are all combined at the end of the section.

\runinhead{Emission angles.} 
Since emission angles are periodic, a Gaussian aleatoric uncertainty is inappropriate.
Instead of the Gaussian negative log-likelihood (NLL) used in \S\ref{sec:deepens}, the von Mises (VM) distribution NLL is a better loss function choice. This is the maximum entropy distribution for circular data with specified expectation. For a random 2D unit vector $\mathbf{x}$
\begin{equation}
    {\rm VM}(\mathbf{\mu},\kappa) \equiv p(\mathbf{x}|\mathbf{\mu},\kappa) = \frac{\exp (\kappa \mathbf{\mu}^T\mathbf{x})}{2\pi I_0(\kappa)}.
\end{equation}
where $I_0$ is the modified Bessel function of the first kind.
This can also be considered a 1D distribution over the polar angle $\theta$ of vector $\mathbf{x}$:
\begin{equation}
    {\rm VM}(\theta_{\mu},\kappa) \equiv p(\theta|\theta_{\mu},\kappa) = \frac{\exp (\kappa \cos(\theta - \theta_{\mu}))}{2\pi I_0(\kappa)}.
\end{equation}
The VM distribution is parameterized by concentration parameter $\kappa$; for large $\kappa$ the VM converges to a Gaussian with variance $\sigma^2 = 1/\kappa$. For small $\kappa$ the VM converges to a uniform distribution.
This more appropriately reflects the distribution of predictions $\hat{\theta}$, which are clearly periodic.

Predicting scalar periodic values directly is tricky for standard DNNs, so the emission angle $\theta$ can be parameterized as a 2D vector $\mathbf{v} = (\rm{cos}2\theta,\rm{sin}2\theta)$. Now the DNNs attempt to predict the 2D vector $\hat{\mathbf{v}}$ \cite{peirson_deep_2021}. Only $-\pi/2 \leq \hat{\theta} < \pi/2$, as opposed to $-\pi \leq \hat{\theta} < \pi$, are required for polarization estimation since the EVPA $-\pi/2 \leq \theta_0 < \pi/2$, i.e. polarization is a quadrupolar signal. However, the full $2\pi$ emission angle can be useful when dealing with systematics and extended source analysis. If predicting the full $2\pi$ emission angle is necessary, an additional loss term for $\mathbf{v}_2 = (\rm{cos}\theta,\rm{sin}\theta)$ can be included \cite{peirson_deep_2021}. Predicting only $\mathbf{v}_2$ gives very poor results for low energy tracks because of direction ambiguity, and should be avoided.

Each DNN model in the ensemble computes the emission angle loss for track image $\mathbf{x}$ with true direction $\theta$ as the VM NLL 
\begin{equation}
    L_{\theta}(\mathbf{v} \mid \mathbf{x}) = -\hat{\kappa}^a(\hat{\mathbf{v}}.\mathbf{v}) + \log I_0(\hat{\kappa}^a) .
    \label{eqn:loss}
\end{equation}
where $\hat{\kappa}^a$ are the DNN predicted aleatoric VM uncertainties.
The epistemic uncertainties can also be assumed to follow ${\rm VM}(0,\kappa^e)$; $\kappa^e$ can be estimated from the output of a deep ensemble with M DNNs $\{\hat{\theta}_{j}\}^M_{j=1}$ using the appropriate maximum likelihood estimator:
\begin{equation}
    \bar{R}^2 = \left(\frac{1}{N}\sum_{j=1}^M\cos2\hat{\theta}_{j}\right)^2 + \left(\frac{1}{N}\sum_{j=1}^M\sin2\hat{\theta}_{j}\right)^2 
\end{equation}
\begin{equation}
    \label{eqn:epis}
    \frac{I_1(\hat{\kappa}^e)}{I_0(\hat{\kappa}^e) } = \bar{R},    
\end{equation}
The total VM predictive error on each track angle $\hat{\theta}$ is approximated by summing the aleatoric and epistemic variances, as in \S\ref{sec:deepens}: 
\begin{equation}
\frac{1}{\hat{\kappa}} =   \frac{1}{\hat{\kappa}^e} + \frac{1}{M}\sum_{j=1}^M \frac{1}{\hat{\kappa}^a_{j}}
\end{equation}

\runinhead{Absorption points.} To predict the (x,y) photon absorption point coordinates given a track image $\mathbf{x}$, a simple Gaussian NLL loss function suffices:
\begin{equation}
     \label{eqn:DE_abs_loss}
    L_{\rm abs}(x,y \mid \mathbf{x}) = \frac{1}{2}\log\hat{\sigma}_{\rm abs}^2 + \frac{1}{2\hat{\sigma}_{\rm abs}}\bigg\|
    \begin{bmatrix}
    x \\ y
    \end{bmatrix}
    - \begin{bmatrix}
    \hat{x} \\ \hat{y}
    \end{bmatrix}\bigg\|^2.
\end{equation}
As in eq.\ref{eqn:log_loss}, the DNN actually predicts the log variance for numerical stability. If desired, the epistemic error for the absorption point predictions can be calculated in the same way as \S\ref{sec:deepens}.

\runinhead{Photon energy.}
Similarly to the absorption point loss eq.\ref{eqn:DE_abs_loss}, the energy loss can be stated as a simple MSE with the aleatoric error included:
\begin{equation}
     \label{eqn:Eloss}
    L_{E}(E \mid \mathbf{x}) = \frac{1}{2}\log\hat{\sigma}_E^2 + \frac{1}{2\hat{\sigma}_E}\big\|
    E
    - \hat{E}\big\|^2,
\end{equation}
and like the absorption points, the epistemic error can be calculated over the deep ensemble as in \S\ref{sec:deepens}.

Predicting the energy, while simple at first sight, is more tricky than absorption points because of tail tracks.
The presence of tail tracks, \S\ref{sec:tail}, can cause the DNNs to produce biased energy estimates $\hat{E}$ with this loss function. Tail tracks look like lower energy tracks than their true energy label suggests, and this can lead to two possibilities:
\begin{itemize}
    \item If tail tracks are difficult to distinguish from peak tracks this will cause the DNN to skew its overall predictions upward in energy to minimize eq.\ref{eqn:Eloss}.
    \item If low energy tail tracks are difficult to distinguish from high energy tail tracks, then most tail tracks will be assigned the mean energy of the training distribution as this minimizes the expected value of eq.\ref{eqn:Eloss}. This is usually undesirable for astrophysical spectra.
\end{itemize}
In principle these two effects could be remedied in a number of ways, for example by adjusting the loss function eq.\ref{eqn:Eloss} \cite{peirson_deep_2021} or by a careful look at the predicted energy uncertainty (tail tracks should have much higher uncertainties).
In practice, since tail tracks also affect emission angle reconstruction and by extension polarization estimation, the recommended approach is to remove tail tracks from the analysis to the greatest extent possible \cite{peirson_towards_2021}. This works well only if tail tracks are distinguishable from peak tracks. For IXPE GPD events this is fortunately the case, but a different approach may be better for other imaging X-ray polarimeters. In \S\ref{sec:removetail} we describe a DNN approach to identify and remove tail tracks. \\

The full DNN loss function to be minimized is now
\begin{equation}
\label{eqn:fullloss}
    L(\theta, x, y, E \mid  \mathbf{x}) = L_{\theta} + L_{\rm abs} + L_{E} + \alpha \|\mathbf{w}\|^2,
\end{equation}
where the final term is an L2 regularization on the DNN weights controlled by hyperparameter $\alpha$, to prevent overfitting the training data. The predicted aleatoric uncertainties act as automatically tuned hyperparameters for the remaining loss terms \cite{kendall_multi-task_2018}. Given a track image input $\mathbf{x}$, an individual DNN outputs an 8 dimensional vector: $(\hat{\mathbf{v}}, \hat{\kappa}^a,\hat{x}, \hat{y}, \hat{\sigma}_{\rm abs}, \hat{E}, \hat{\sigma}_{E})$.

\subsection{Removing tail tracks}
\label{sec:removetail}
Photoelectron tracks from events converting outside the main detector gas are known as tail tracks, \S\ref{sec:tail}. For both DNNs and classical methods they cause problems with photon energy prediction, discussed in \S\ref{sec:deepset}. Furthermore, tail tracks have comparatively high emission angle prediction error (low polarization sensitivity \S\ref{sec:tail}) and \S\ref{sec:perform} will show that this error is not properly captured by deep ensemble predictive uncertainties. One possible approach to remedy the tail track issue is to attempt to excise tail tracks from the data analysis entirely.     

\begin{figure}[t]
\centering
\includegraphics[width=1.05\textwidth]{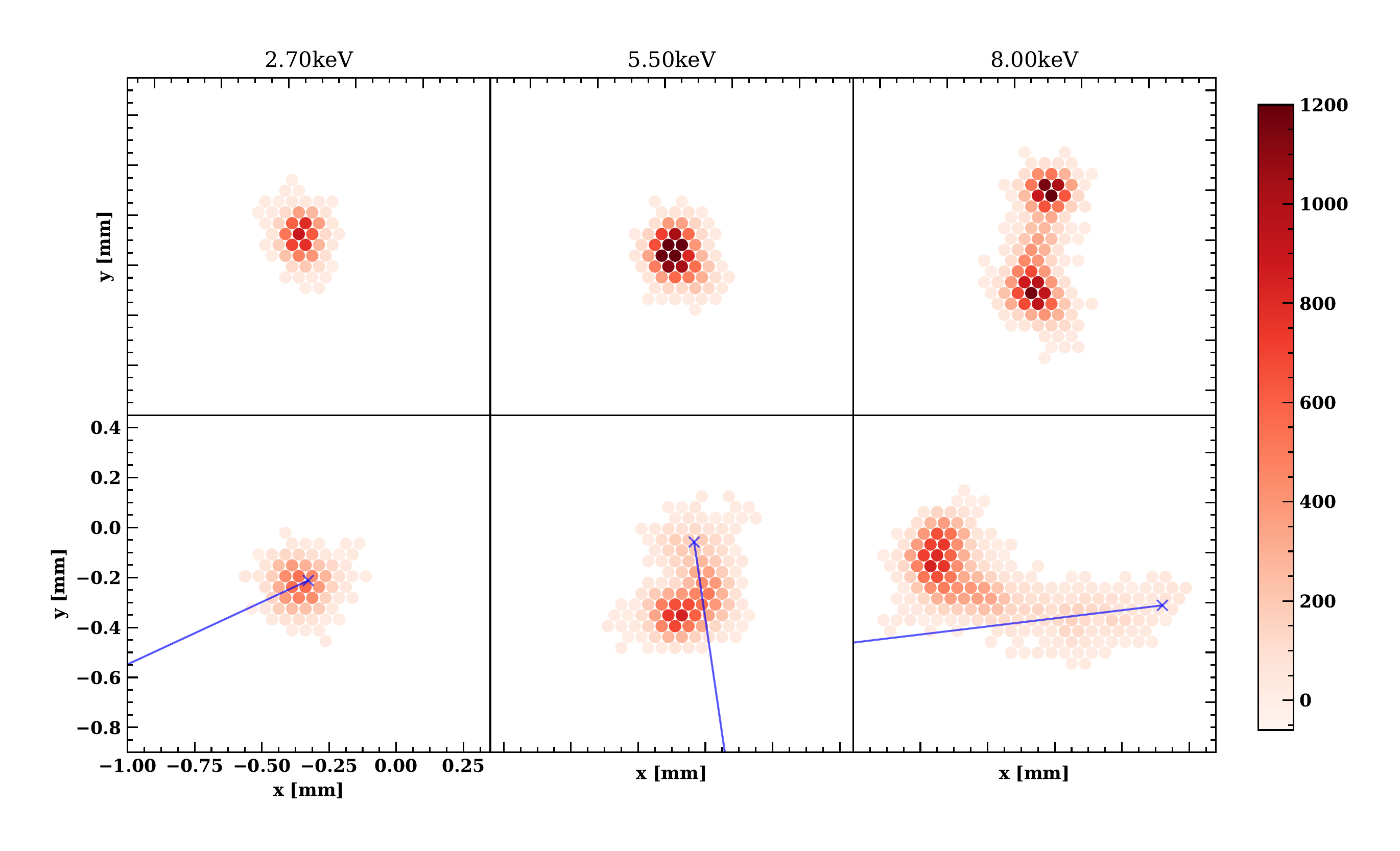}
\caption{Example tail events (top row) and peak events (bottom row) for three different recovered energies. All plots follow the same spatial and color scale. These tail events convert in the GEM. Color denotes charge deposited in a given detector pixel. Note that tail events are more compact, short and with high charge density at maximum.}
\label{fig:tailvpeak}
\end{figure}

For IXPE's GPDs, tail tracks differ in morphology from peak events converting in the gas. Fig.~\ref{fig:tailvpeak} compares events with the same linearly recovered energy (largely determined through the summed pixel counts, the summed energy deposition in the gas). Since the GEM and window material have a lower mean free path for photoelectrons, detected tail events are typically due to photo-electrons ejected close to the GEM/window normal. Their tracks are thus more compact, with higher counts/pixel for the same recovered energy. Window events have larger drift diffusion than GEM conversions. Distinguishing peak and tail tracks can be formed as a computer vision classification problem, so a DNN would be a good model choice and could be trained on simulated events to recognize the differences. 

A good approach for peak vs. tail track classification would be to include the classification task into the existing track reconstruction deep ensemble, \S\ref{sec:deepset}. This would likely improve performance over training a separate model and at first glance would only require an additional loss term in eq.\ref{eqn:fullloss}. However, this approach would not fix the problem of improper energy predictions and emission angle uncertainties unless more complex cross terms to eq.\ref{eqn:fullloss} were also added. The cross terms would ensure, for example, that tracks likely to be tail events have higher energy uncertainty. Opting for a simpler approach, \cite{peirson_towards_2021} trained a separate peak vs. tail classifier DNN and used only peak tracks to train the track reconstruction deep ensemble. The \cite{peirson_towards_2021} separation approach is described in this section, but a unified deep ensemble approach to tail vs peak classification would likely work better and is discussed as a future direction in \S\ref{sec:6}.

\begin{figure}[t]
\sidecaption
\includegraphics[scale=.65]{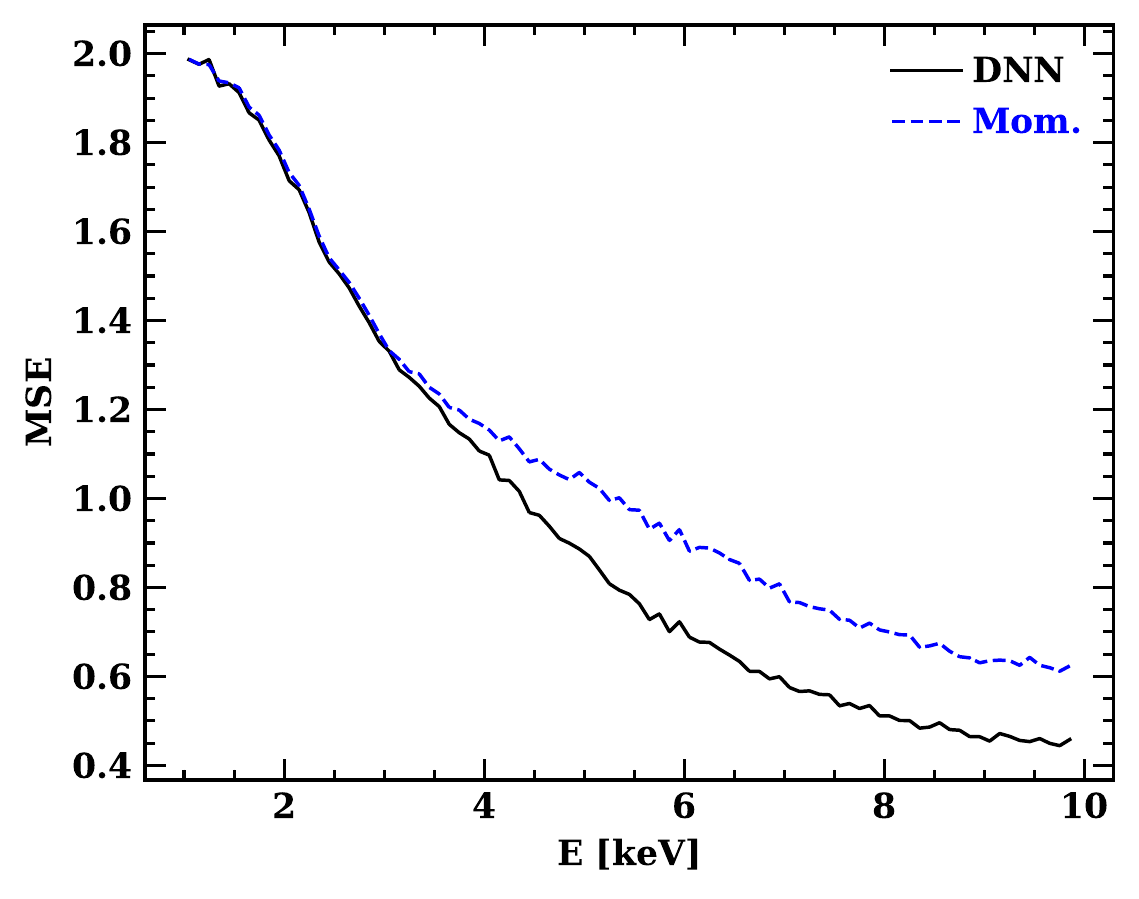}
\caption{Mean squared error on emission angle prediction as a function of photon energy, measured using eq.~\ref{eqn:vmse}, for the classical moment analysis and deep ensemble.}
\label{fig:mse}      
\end{figure}

Setting up a peak vs. tail DNN classification model is the same as setting up a multitask deep ensemble \S\ref{sec:deepset}, only now the loss function to be minimized is different. In binary classification tasks, DNNs minimize the cross-entropy loss, eq.\ref{eqn:eglossbin}.
Given a single input, a DNN outputs a single scalar between 0 and 1 that represents the probability an input is the first class. \cite{peirson_towards_2021} use exactly the same ResNet-18 architecture and data preprocessing, \S\ref{sec:data}--\ref{sec:geom}, when training the peak vs. tail classifier DNN; the only notable differences are the loss function and data labels.

\subsection{Training and ensemble selection}
\label{sec:trainens}
Individual DNN training procedures should follow the guidelines in \S\ref{sec:training}, \S\ref{sec:val}. Specifics will depend on the particular DNN architecture chosen. CNNs for computer vision problems train particularly well using stochastic gradient descent with momentum \cite{sutskever_importance_2013} as the optimization algorithm. Normalizing the inputs and outputs before training typically helps improve convergence. From \cite{peirson_deep_2021}:
\begin{quotation}
The ResNet-18 architecture trains in a reasonable amount of time ($\sim 15$ hours for 150 epochs on 4 Nvidia Titan GPUs, using a batch size of 2048). 
Before training, we normalize the track images, subtracting the pixel-wise mean from each track image and dividing by the pixel-wise standard deviation (where the mean and standard deviation are calculated over the full training set). The track energy and absorption point labels are similarly processed.
Normalizing the training data helps prevent vanishing and exploding gradients during the NN training procedure and lead to faster convergence. We use stochastic gradient descent with momentum as our optimizing algorithm, typical in computer vision tasks \cite{sutskever_importance_2013}, with a stepped decaying learning rate starting at 0.01.
We choose batch sizes of 512, 1024, 2048 tracks ... We tune the hyperparameters to minimize the MSE for the validation set.
\end{quotation}
Once multiple DNNs are trained with optimized hyperparameters, individual DNNs should be chosen at random to compose the ensemble. This ensures accurate epistemic uncertainty estimates.

\begin{figure}[t]
\centering
\includegraphics[width=1.0\textwidth]{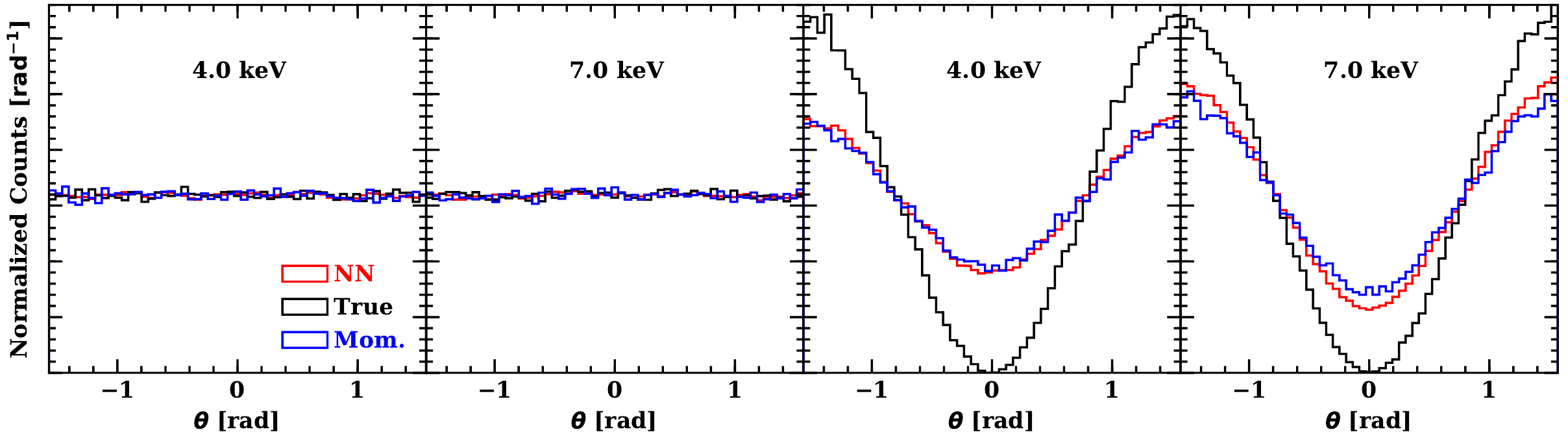}
\caption{Emission angle recovery for unpolarized, $p_0 = 0$, (left two panels) and 100\% polarized, $p_0 = 1$, (right two panels) simulated IXPE data for $4.0$ and $7.0$\,keV. The true photoelectron angle distribution is shown in black; standard moment analysis reconstruction is in blue and deep ensemble in red.}
\label{fig:angledist}
\end{figure}

\subsection{Performance}
\label{sec:perform}
The performance of deep ensembles in track reconstruction is compared to the classical moment analysis \S\ref{sec:2} on simulated IXPE data. The results shown are for peak events only unless otherwise specified.

\runinhead{Emission angles.}  Two important measures of performance should be evaluated for emission angle recovery: the accuracy of recovery and the quality of uncertainty estimates. The accuracy of emission angle recovery can be measured, for example, by the MSE on a test dataset
\begin{equation}
\label{eqn:vmse}
    \frac{1}{N}\sum^N_{i=1} \|\mathbf{v}_i - \mathbf{\hat{v}}_i\|^2.
\end{equation}
Fig.~\ref{fig:mse} gives the above MSE as a function of the true track energy on an unpolarized IXPE test dataset for both classical moment analysis and deep ensemble approaches. At low energies, deep ensembles fail to improve significantly over the moment analysis; when resolution is low, track image moments contain all the relevant information. 

\begin{figure}
\centering
\includegraphics[width=1.0\textwidth]{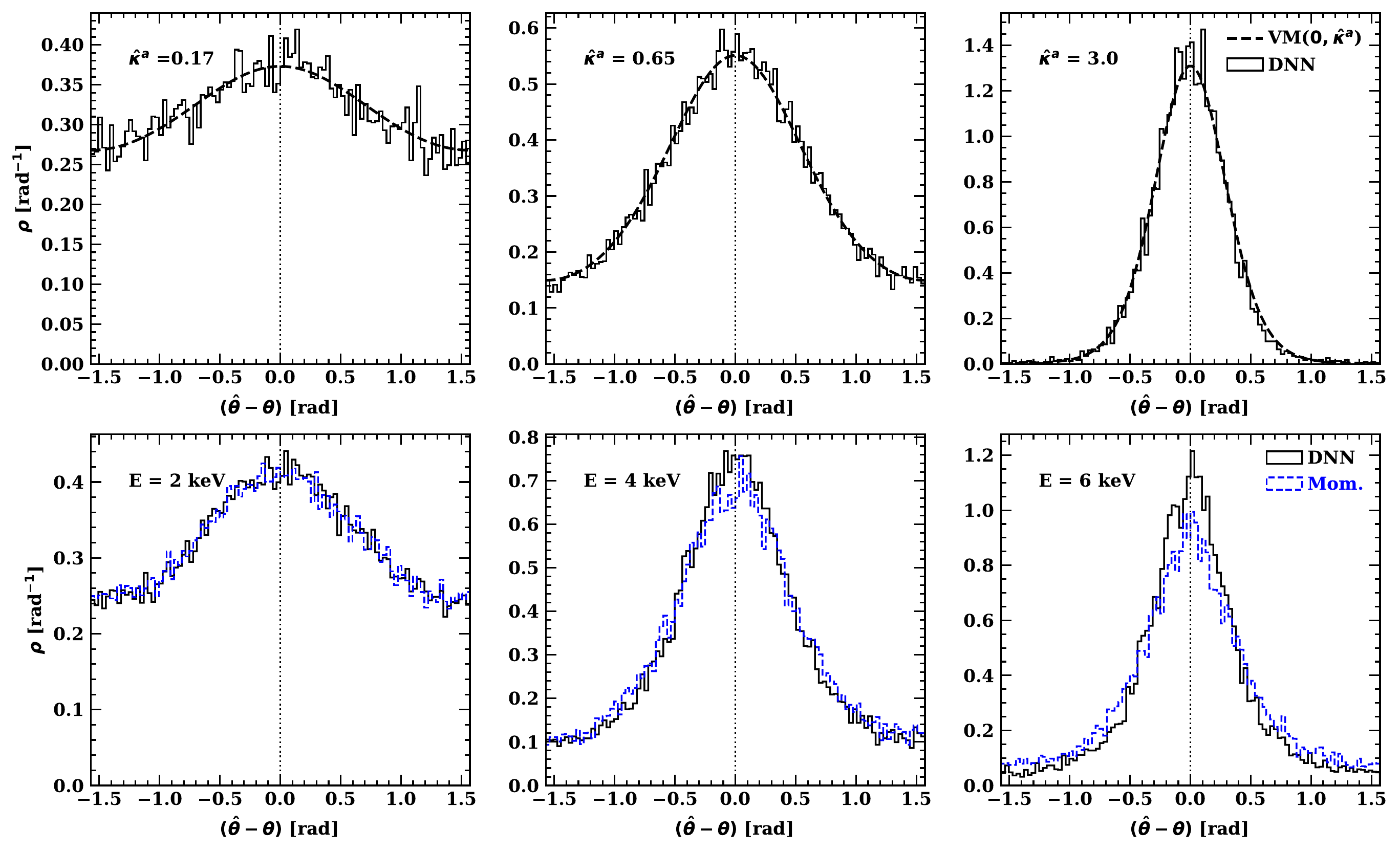}
\caption{\textit{Top:} Histograms of the DNN emission angle residual for three different DNN predicted aleatoric VM concentration parameters $\hat{\kappa}^{a}$. The DNN predicted VM error distributions are over-plotted. The DNN is able to correctly predict its emission angle aleatoric error distribution. \textit{Bottom:} Histograms of emission angle residuals at three different energies for DNNs and the moment analysis. All methods are unbiased (centered at zero) and at higher energies the DNNs have better recovery of the true emission angle $\theta$, as evidenced by the larger peak heights.}
\label{fig:kappa}
\end{figure}

It is also essential to confirm that the estimates $\hat{\theta}$ are unbiased.
Fig.\ref{fig:angledist} gives the distribution of emission angles $\hat{\theta}$ recovered by the deep ensemble and moment analysis for an unpolarized and polarized source at various energies. For both methods, emission angle estimates show no systematic biases, and at higher energies for the polarized source the improved signal recovery of the deep ensemble is visible. 
As an additional check, the bottom row of fig.~\ref{fig:kappa} displays the distribution of $\hat{\theta} - \theta$ for the moment analysis and deep ensemble at three different true energies. Both the moment analysis and deep ensemble have symmetric error distributions centered at zero for all energies. At higher energies, the lower dispersion and thus higher accuracy of the deep ensemble is again clear.

\begin{figure}
\centering
\includegraphics[width=1.05\textwidth]{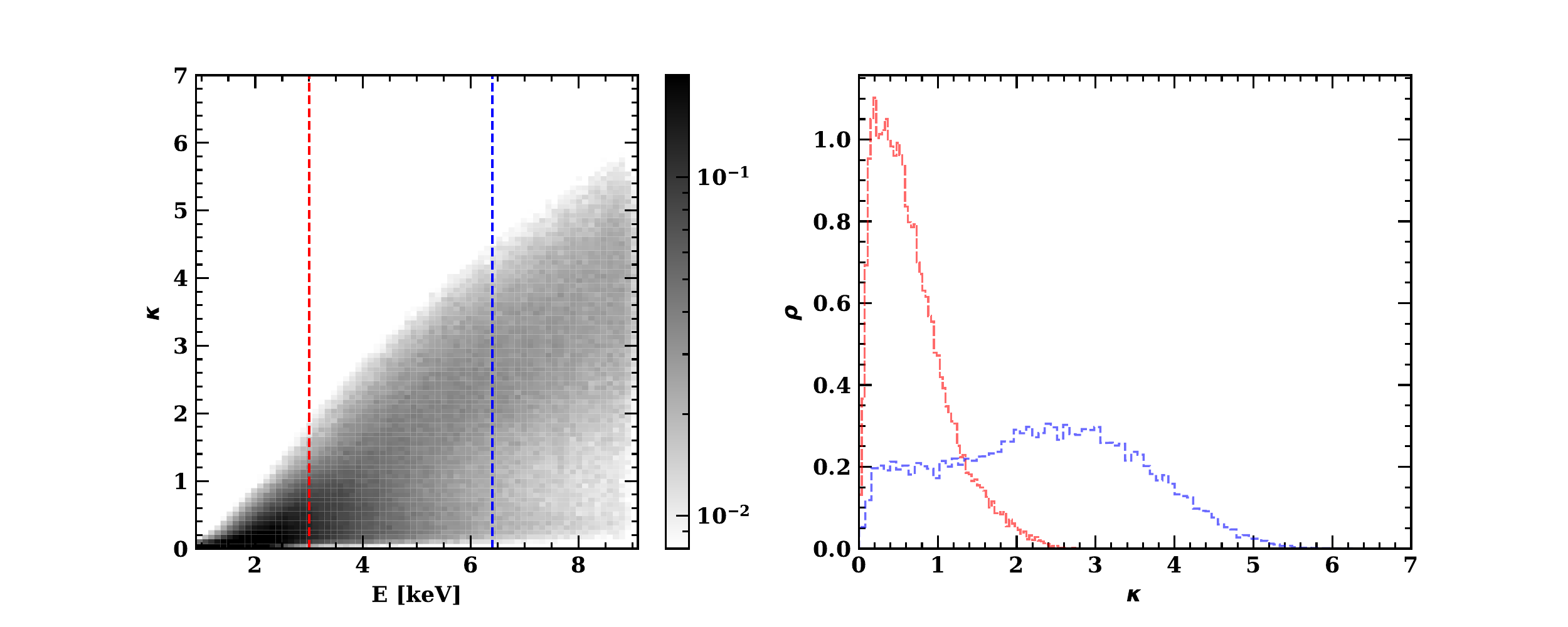}
\caption{Distribution of deep ensemble predicted concentration parameters $\hat{\kappa}$ across the IXPE energy spectrum (left). The right-hand plot shows the $\hat{\kappa}$ distribution for two specific energies $3.0$ keV (red), $6.4$ keV (blue). High concentration parameter $\kappa$ means a low predictive uncertainty.}
\label{fig:kappadist}
\end{figure}

A simple way to assess the deep ensemble predicted uncertainty estimates $\hat{\kappa}$ is to consider the distribution of $\hat{\theta} - \theta$ for fixed $\hat{\kappa}$. According to the assumptions in \S\ref{sec:deepset}, $\hat{\theta} - \theta$ should follow a VM(0,$\hat{\kappa}$) if the DNNs have been properly trained. In the top row of fig.~\ref{fig:kappa}, the distributions of $\hat{\theta} - \theta$ for three different fixed $\hat{\kappa}$ are displayed with VM(0,$\hat{\kappa}$) overlayed. In all cases, the distributions match very closely. Epistemic errors are negligible for the vast majority of peak tracks, suggesting good model choice and training convergence.

It is also instructive to visualize the distribution of deep ensemble predicted uncertainties as a function of energy, fig.~\ref{fig:kappadist}. For higher energies a wider range of concentration parameters are predicted and the $\hat{\kappa}$ peak moves progressively higher. Below 2keV, barely any polarization signal is recoverable with IXPE. Fig.~\ref{fig:kappadist} highlights the strong heteroskedasticity in imaging X-ray polarimetry.

\begin{figure}[b]
\centering
\includegraphics[width=1.0\textwidth]{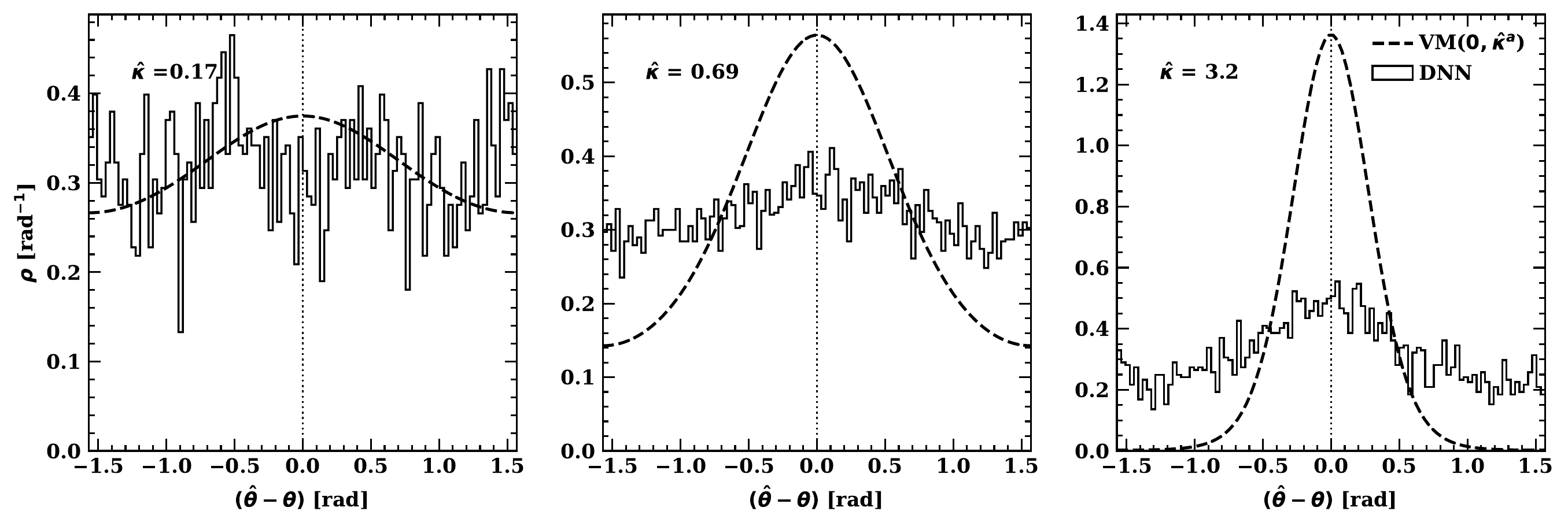}
\caption{Histograms of the emission angle residual at three different DNN predicted concentration parameters $\hat{\kappa}$ for tail events only. The predicted DNN error distributions are overplotted. When tail events are treated as if they were normal events, the DNNs do not learn to predict the appropriate uncertainties.}
\label{fig:kappa_tail}
\end{figure}

For a deep ensemble trained on both peak and tail tracks, the predicted uncertainties for tail tracks are untrustworthy. Comparing the error distributions in fig.~\ref{fig:kappa_tail} to \ref{fig:kappa}, tail track error is not captured by deep ensemble uncertainty predictions, even with the epistemic uncertainty included. Furthermore, at low energies, tail track emission angle estimates are potentially biased.

\begin{figure}[t]
\centering
\includegraphics[width=1.0\textwidth]{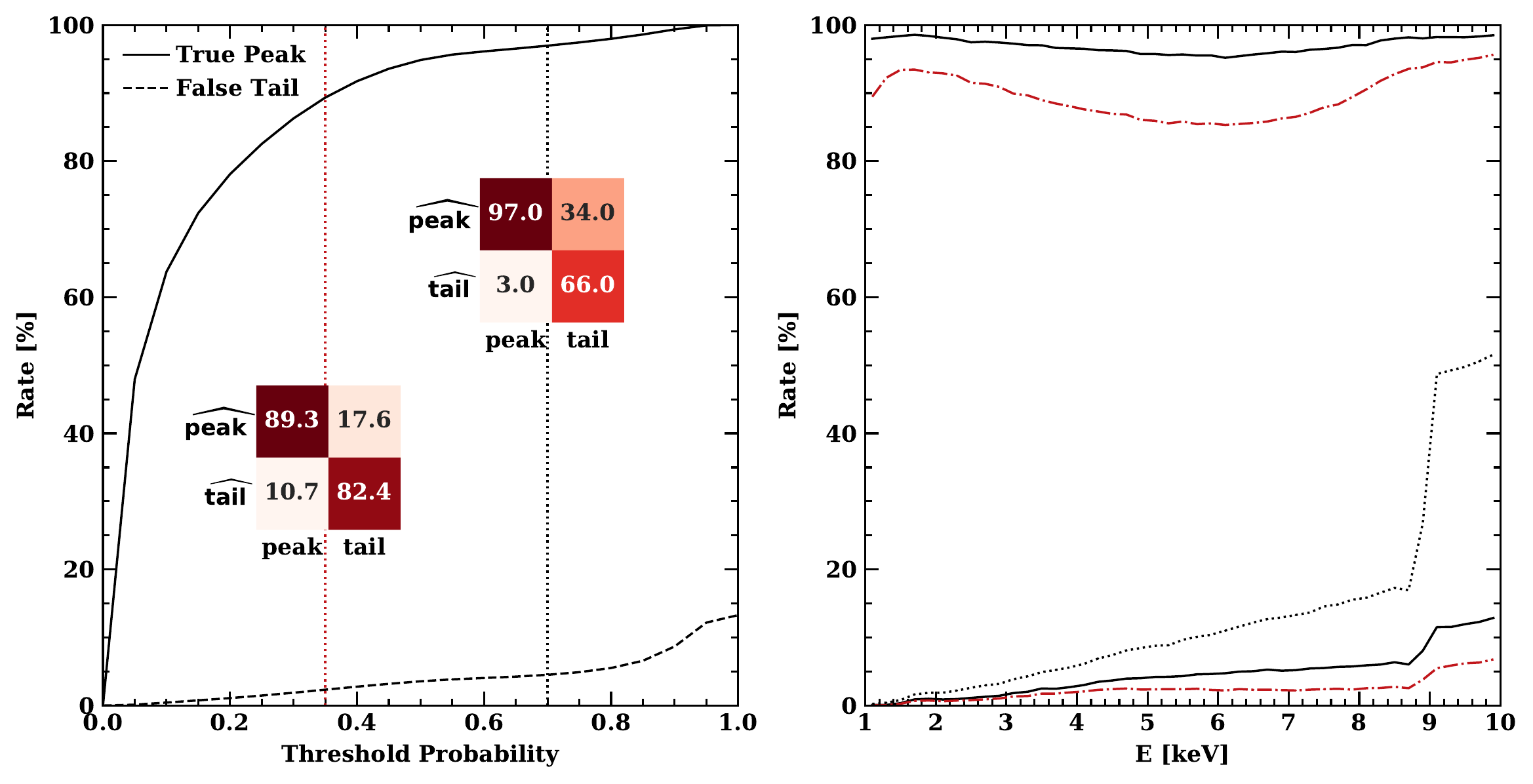}
\caption{\textit{Left:} The solid curve shows the fraction of peak events retained as a function of the tail probability cut, while the dashed curve shows the fraction of the cut sample from the remaining tail events. Insets show the confusion matrices, normalized by column, for our adopted 70\% cut and a 35\% cut. \textit{Right:} The top curves show the peak retention as a function of energy (black solid -- 70\% cut, red dot-dash -- 35\% cut). Below, the dotted curve shows the uncut fraction of the sample due to tail events, while the lower black and red curves show the residual tail pollution (70\% and 35\% cut, respectively). Depending on how harmful tail events are to the desired measurements, different cut levels could be appropriate.}
\label{fig:tailcut}
\end{figure}

\runinhead{Tail vs. peak classification.} For tail vs. peak classification, there is no equivalent classical method to compare the DNN performance, so this section lists the classification results achieved by \cite{peirson_towards_2021}. They use the predicted tail probability as an event threshold cut. Fig.~\ref{fig:tailcut} shows the effectiveness of different tail probability cuts. If all events with a DNN predicted tail probability higher than 70\% are removed, a loss of only 3\% of true peak events is incurred while 66\% of true tail events are successfully removed. Using a DNN classifier, only a small loss to signal events need be incurred to remove the majority of harmful tail events. Fig.~\ref{fig:tailcut} also shows how remaining tail and peak events are distributed in terms of true energy after a tail probability threshold cut. Remaining tail events come from all energies in proportion to their original population, while misidentified peak events come from middle energies where peak and tail events overlap most.

At least for IXPE GPDs, it is certainly possible to meaningfully distinguish between tail and peak events. 
The upcoming section on photon energy recovery and \S\ref{sec:6} demonstrate how tail cuts based on DNN probabilities can be used to improve energy and polarization resolution.

\begin{figure}[t]
\sidecaption
\includegraphics[scale=.65]{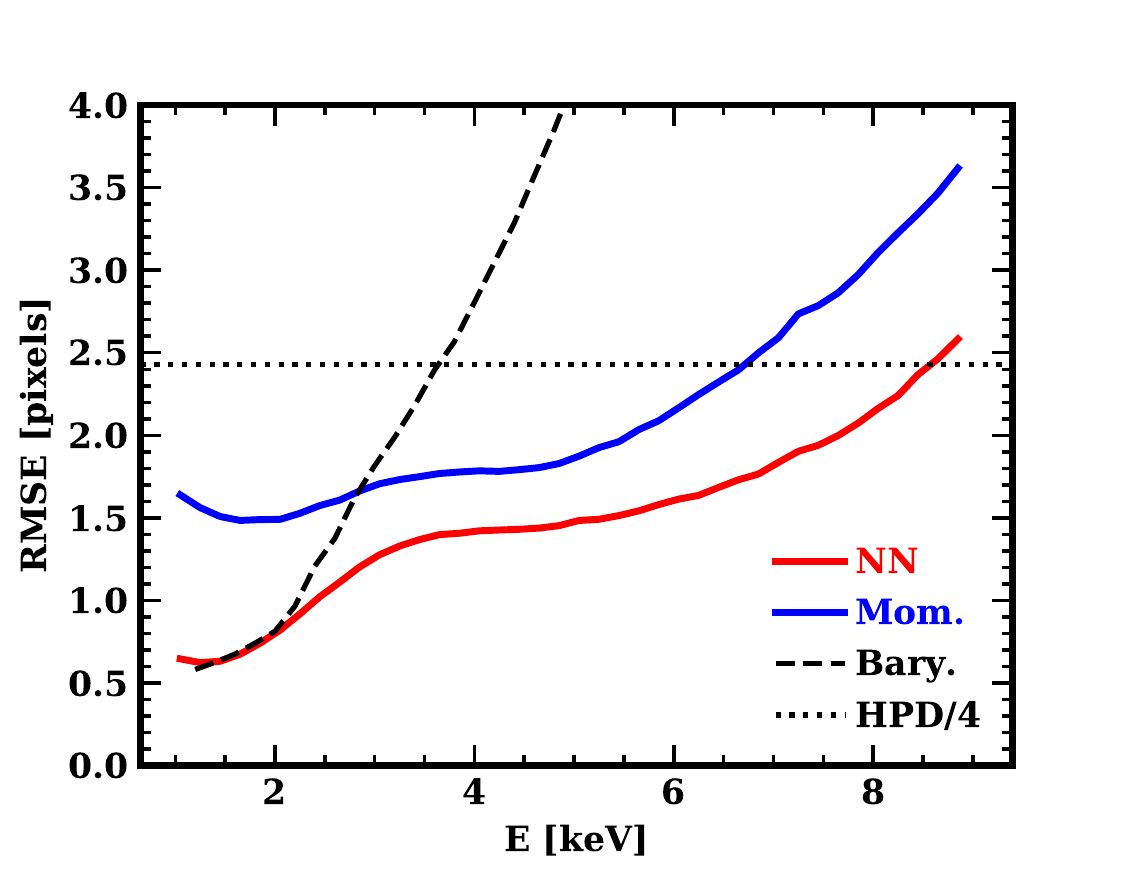}
\caption{Photon absorption point localization accuracy using the root mean squared error. The deep ensemble predictions (red) do appreciably better than the moment analysis and matches a barycenter estimate at the lowest energies. For IXPE, all methods are adequate, as localization is much better than the size of the point source image produced by IXPE's mirrors. One quarter of IXPE's half power diameter is illustrated (black dotted line).}
\label{fig:abspts}       
\end{figure}

\runinhead{Absorption points.} 
Absorption point accuracy can be evaluated using the MSE. In fig.~\ref{fig:abspts} the classical moment analysis and deep ensemble predictions are compared at different photon energies. For IXPE, the track barycenter provides a better prediction of the absorption point for very low energies -- the deep ensemble recovers this transition while the moment analysis does not. 

For current generation X-ray polarimeters like IXPE, the telescope PSF is much larger than any absorption point discrepancy, so absorption point accuracy is less essential than emission angle or photon energy accuracy. However, future missions with higher spatial resolution may require very accurate absorption point estimates for detailed imaging of extended sources.

\begin{figure}[t]
\centering
\includegraphics[width=1.0\textwidth]{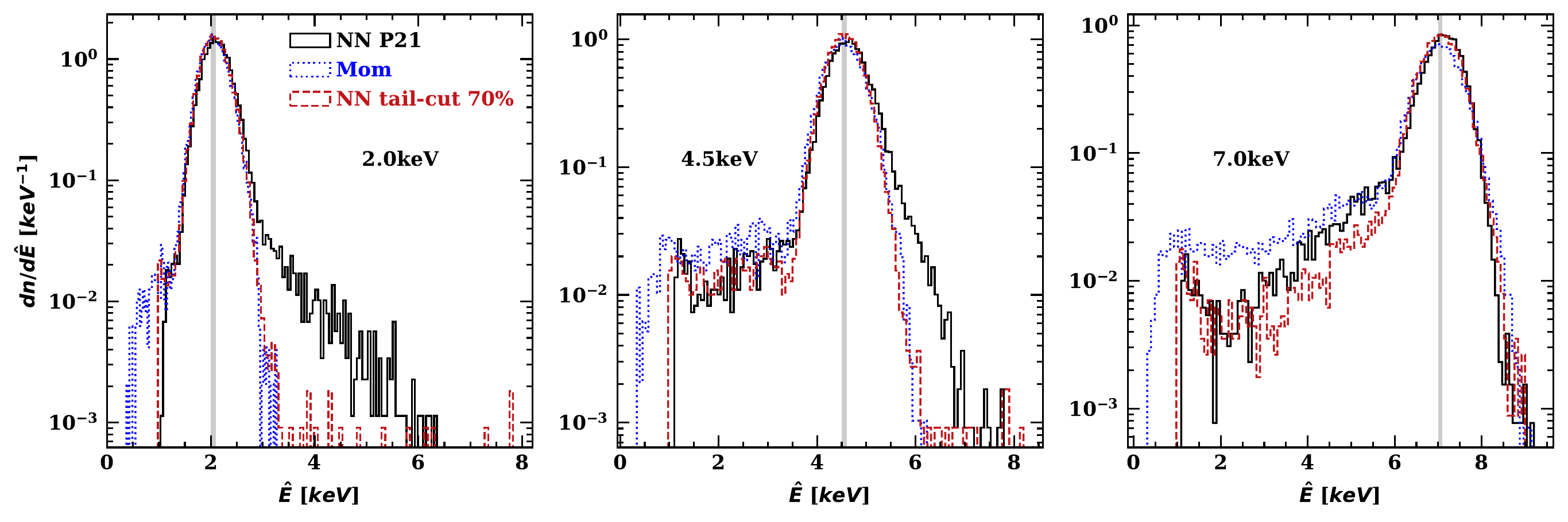}
\caption{Response for three true energies. Note that while the P21 analysis suppressed the large tail rate seen in the Moments processing, events leaked to a high energy tail for medium to low true energies. Our morphological tail cut further suppresses the GEM/window tail, avoids the high energy leakage and achieves comparable or better peak width.}
\label{fig:energy_hist}
\end{figure}

\runinhead{Photon energy.} Energy recovery can be adversely affected by tail events. While the MSE in predicted energy is lower for deep ensemble methods compared to classical linear charge recovery, these results should be treated with caution. Much of this improvement is specious, coming from a mistreatment of tail events.
Fig.\ref{fig:energy_hist} plots the IXPE recovered energy histograms of the classical linear method (blue), a deep ensemble trained including tail events (black) and a deep ensemble trained on only peak events with tail events removed at a 70\% peak vs. tail classifier probability (red). Around the true energy peak all methods perform similarly, with a slight advantage for the DNN based methods. As the energy increases, all methods show a low energy tail in reconstructed energy which is produced by tail events. The standard deep ensemble that does not differentiate peak and tail tracks also shows a high energy tail in reconstructed energy. This tail arises from low energy peak events 'looking' like higher energy tail events, so pushing some events up to higher energy tends to minimize the MSE. As discussed in \S\ref{sec:deepset}, high energy tails in reconstructed energy are undesirable for astrophysical spectra. Training a deep ensemble with only peak events prevents this high-energy tail behavior.

Cutting events based on the predictions of the peak vs. tail classifier successfully reduces tails in the recovered energy histograms. A tail event cut improves the energy resolution at the cost of potentially losing some peak events and thus some polarization signal. \S\ref{sec:perform2} will discuss and quantify the polarization signal loss due to tail event cuts. 

\begin{figure}[t]
\centering
\includegraphics[width=1.0\textwidth]{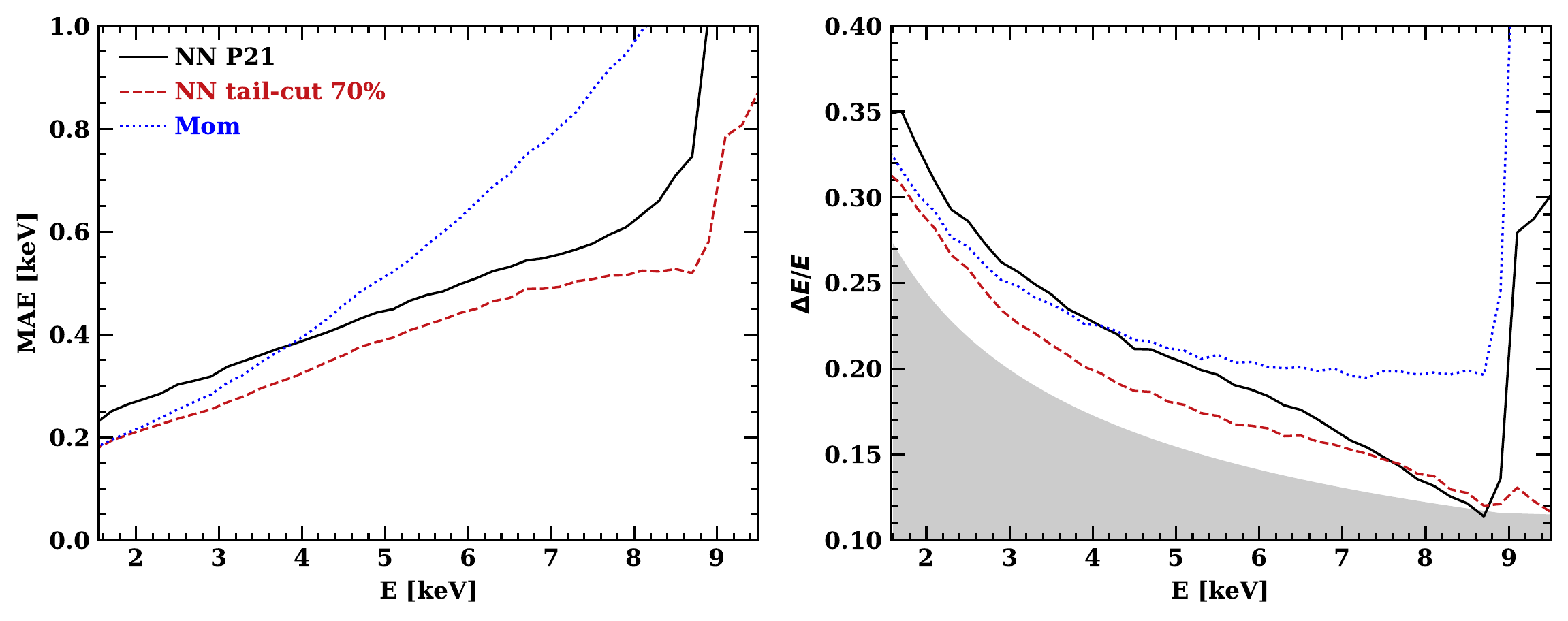}
\caption{\textit{Left:} Mean absolute error in predicted energy as a function of true energy. \textit{Right:} FWHM (3.46$\times$ Median Absolute Deviation) of predicted energy distribution for a source at energy E. The grey band marks the limiting energy resolution in the purely-exponential multiplication regime. NNs with 70\% tail cut performs better on every metric, more accurate per track and tighter resolution at all energies. All methods suffer from the large increase of tail events above 9\,keV.}
\label{fig:energy_mse}
\end{figure}

The energy prediction accuracies (mean absolute error and FWHM) for all three methods as a function of photon energy are given in fig.~\ref{fig:energy_mse}. Removing tail events provides a significant improvement to the energy prediction accuracy.

\newpage
\section{Neural networks for polarization estimation}
\label{sec:5}

\S\ref{sec:polest} explained how to best estimate polarization parameters $(p_0, \theta_0)$ given a group of reconstructed photoelecton emission angles $\{\hat{\theta}_i\}^N_{i=1}$. This section explains how to best estimate polarization parameters given a group of reconstructed photoelecton emission angles \textit{and their associated uncertainties} $\{\hat{\theta}_i, \hat{\kappa}_i\}^N_{i=1}$. These uncertainties can vary for different individual emission angles; the measurements are heteroskedastic.
Emission angles with larger uncertainties should contribute less to the final polarization parameter estimates. In theory, this section need not use DNNs and the methods described can be used for any track reconstruction method that provides (trustworthy) emission angle uncertainties; the uncertainties can be parameterized by any appropriate distribution. In practice, the predicted emission angle uncertainties come from deep ensembles and the uncertainties are assumed to follow a von Mises family of distributions. 

\subsection{Modulation factor}
To develop a polarization estimation method that incorporates emission angle uncertainties, it is essential to specify how these uncertainties affect the recovered emission angle distribution eq.~\ref{eqn:likelihood}. For a source with polarization parameters $(p_0, \theta_0)$, the true photoelectron emission angles follow the distribution eq.~\ref{eqn:likelihood}, while the recovered or measured emission angles follow eq.~\ref{eqn:likelihoodmu}. The difference between the two is captured by the modulation factor $\mu$. The modulation factor summarizes the effect of uncertainty in individual emission angle measurements.  

Consider a single emission angle measurement $\hat{\theta}_i$. Since track reconstruction methods are imperfect and contain sources of error (even DNN based ones as was shown in \S\ref{sec:4}), $\hat{\theta}_i$ can be considered a random variable:
\begin{equation}
    \hat{\theta}_i = \theta_i + \epsilon_i
\end{equation}
where $\theta$ is the true emission angle, which follows the distribution eq.~\ref{eqn:likelihood}, and the measurement error $\epsilon_i$ is a random variable with  
\begin{equation}
\label{eqn:properties}
    \mathbb{E}[\epsilon_i] = 0, {\rm Var}[\epsilon_i] = \sigma^2_i, p(\epsilon_i=0) = p(\epsilon_i = 2\pi) 
\end{equation}
This assumes the $\hat{\theta}_i$ estimate is unbiased; reasonable given both the deep ensemble and moment analysis in \S\ref{sec:perform} are for the most part unbiased estimators. 
The specific distribution for $\epsilon_i$ will depend on the track reconstruction method, but all distributions should follow the properties in eq.~\ref{eqn:properties}: be unbiased, have a finite variance, and be periodic, since $\hat{\theta}$ itself is periodic. 

For any $\epsilon_i$ distribution with the above properties, it is possible to find the distribution for $\hat{\theta}_i$ as the convolution of the $\theta_i, \epsilon_i$ distributions:
\begin{equation}
    p(\hat{\theta}_i\mid p_0,\theta_0,\sigma_i^2) = \frac{1}{2\pi} \big(1 + \mu_ip_0\cos[2(\hat{\theta}_i - \theta_0)] \big),
    \label{eqn:individual}
\end{equation}
where $0 \leq \mu_i \leq 1$, $\mu_i(\sigma^2_i)$ and $\mu_i(\sigma_i^2 = 0) = 1$.
In other words, the distribution of estimators $\hat{\theta}_i$ are the same as the distribution of the true values $\theta_i$ but with a reduced individual modulation factor $\mu_i(\sigma^2_i)$ that depends on the measurement error $\sigma^2_i$. The measurement noise will blur the sinusoidal modulation signal by a factor $\mu_i$ for the specific event $i$. 

The final distribution for a set of recovered emission angles $\{\hat{\theta}_i\}^N_{i=1}$ can now be simply formulated as an equal mixture model of the distributions of all individual recovered emission angles, eq.\ref{eqn:individual}:
\begin{eqnarray}
    p(\hat{\theta}\mid p_0,\theta_0) =& \frac{1}{N}\sum_{i=1}^N\frac{1}{2\pi} \big(1 + \mu_ip_0\cos[2(\hat{\theta} - \theta_0)] \big),\\
    =& \frac{1}{2\pi} \big(1 + \big( \frac{1}{N}\sum_{i=1}^N \mu_i \big)p_0\cos[2(\hat{\theta} - \theta_0)] \big),\\
    =& \frac{1}{2\pi} \big(1 + \mu p_0\cos[2(\hat{\theta} - \theta_0)] \big).
    \label{eqn:mix}
\end{eqnarray}
Thus, the final modulation factor of a set of emission angles with varying individual measurement errors $\{\hat{\theta}_i, \sigma_i^2\}^N_{i=1}$ is an average over all individual modulation factors 
\begin{equation}
    \mu = \frac{1}{N}\sum_{i=1}^N \mu_i(\sigma_i^2),
\end{equation}
where individual modulation factors each depend on the individual measurement errors $\sigma_i^2$.

For a specific example of an individual modulation factor, consider the case where $\epsilon_i$ is VM(0,$\kappa_i$) between $[0,\pi]$. This is the uncertainty distribution learned by the deep ensembles in \S\ref{sec:4}. 
\begin{equation}
p(\epsilon_i) = \frac{1}{2\pi I_0(\kappa_i)}\exp(\kappa_i\cos2\epsilon_i).
\label{eqn:vm}
\end{equation}
This distribution meets all the assumptions stipulated in eq.~\ref{eqn:properties}. Evaluating the convolution of the $\theta$ (eq.\ref{eqn:likelihood}) and $\epsilon$ (eq.\ref{eqn:vm}) distributions, one finds
\begin{equation}
    p(\hat{\theta}_i\mid p_0, \theta_0, \kappa_i) = \frac{1}{2\pi} \big(1 + \frac{I_1(\kappa_i)}{I_0(\kappa_i)}p_0\cos[2(\hat{\theta}_i - \theta_0)] \big).
    \label{eqn:vm_hat}
\end{equation}
Comparing to eq.~\ref{eqn:individual}
\begin{equation}
    \mu_i = \frac{I_1(\kappa_i)}{I_0(\kappa_i)},
    \label{eqn:weight_func}
\end{equation}
and the modulation factor for set of tracks $\{\hat{\theta}_i, \kappa_i\}^N_{i=1}$ is 
\begin{equation}
    \mu = \frac{1}{N}\sum_{i=1}^N \frac{I_1(\kappa_i)}{I_0(\kappa_i)},
\end{equation}

\subsection{Weighted maximum likelihood estimator}
In \S\ref{sec:polest}, it was assumed when estimating polarization parameters that the modulation factor is a constant, $\mu$. The previous section showed how the modulation factor arises from the measurement errors of many individual emission angle estimates. If individual uncertainties are known, rather than assuming these lead to one constant $\mu$, individual modulation factors can be directly incorporated into the likelihood function for the polarization parameters. 

With individual emission angle measurement errors known, the likelihood eq.~\ref{eqn:likelihood_stoks} can be expressed as a more informative likelihood:
\begin{equation}
    L(\{\hat{\theta}_i\}_{i=1}^N\mid\mathcal{Q},\mathcal{U}) =  \prod_{i=1}^N\frac{1}{2\pi} \big(1 + \mathcal{Q}\mu_i\cos2\hat{\theta}_i + \mathcal{U}\mu_i\sin2\hat{\theta}_i \big).
    \label{eqn:prob_stoks}
\end{equation}
Now each estimated emission angle $\hat{\theta}_i$ comes with its own individual modulation factor $\mu_i$, as opposed to a global $\mu$. This likelihood can be maximized in exactly the same way as eq.~\ref{eqn:likelihood_stoks}. Following \S\ref{sec:polest}, assuming $|\mathcal{Q}\mu_i| << 1$, $|\mathcal{U}\mu_i| << 1$, the maximum likelihood estimators for Stokes' parameters $(\mathcal{Q}, \mathcal{U})$ are now
\begin{equation}
    \hat{\mathcal{Q}} = \frac{2}{\sum_{i=1}^N\mu_i^2} \sum^N_{i=1}\mu_i\cos2\hat{\theta}_i,
    \label{eqn:qestw}
\end{equation}
\begin{equation}
    \hat{\mathcal{U}} = \frac{2}{\sum_{i=1}^N\mu_i^2} \sum^N_{i=1}\mu_i\sin2\hat{\theta}_i.
    \label{eqn:uestw}
\end{equation}
Following the notation in \cite{kislat_analyzing_2015} and \cite{peirson_towards_2021}, one can define the individual event weights
\begin{equation}
    w_i = \mu_i
\end{equation}
and the total modulation factor (now a weighted average)
\begin{equation}
    \mu = \frac{\sum^N_{i=1} w_i\mu_i}{\sum_{i=1}^N w_i} = \frac{\sum^N_{i=1} w_i^2}{\sum_{i=1}^N w_i}.
\end{equation}                                     
Then eqs.\ref{eqn:qestw},\ref{eqn:uestw} are 
\begin{equation}
    \hat{\mathcal{Q}} = \frac{2}{\mu\sum^N_{i=1}w_i} \sum^N_{i=1}w_i\cos2\hat{\theta}_i,
    \label{eqn:qestww}
\end{equation}
\begin{equation}
    \hat{\mathcal{U}} = \frac{2}{\mu\sum^N_{i=1}w_i} \sum^N_{i=1}w_i\sin2\hat{\theta}_i.
    \label{eqn:uestww}
\end{equation}
Now when estimating the polarization parameters, each event is weighted by its expected signal $w_i = \mu_i$. As in \S\ref{sec:polest}, \cite{kislat_analyzing_2015} derive the posterior distribution for estimators eqs.~\ref{eqn:qestww},\ref{eqn:uestww}. The posterior distribution in the weighted case is the same as eq.\ref{eqn:posterior} but now $N$ is replaced with
\begin{equation}
N_{\rm eff} = \frac{\big(\sum_{i=1}^N w_i\big)^2}{\sum_{i=1}^N w_i^2},
\end{equation}
the effective number of events. Because the posterior distribution has changed, now
\begin{equation}
    \label{eqn:mdpw}
    {\rm MDP}_{99} = \frac{4.29}{\mu \sqrt{N_{\rm eff}}}.
\end{equation}
Weighting events reduces the total number of events contributing to a measurement of the polarization parameters, making $\sqrt{N_{\rm eff}}$ smaller, but it increases the modulation factor $\mu$ by more. Overall the MDP$_{99}$ is reduced, improving the polarization measurement. Since $w_i = \mu_i$ is the maximum likelihood estimator, this is the optimal event weighting scheme. It is also possible to arrive at this conclusion in reverse: starting with weighted estimators eq.~\ref{eqn:qestww},\ref{eqn:uestww} and MDP$_{99}$, eq.\ref{eqn:mdpw}, \cite{peirson_towards_2021} prove that $w_i = \mu_i$ minimizes the MDP$_{99}$, or equivalently maximizes the signal-to-noise ratio. 

\begin{figure}[t]
\sidecaption
\includegraphics[scale=.6]{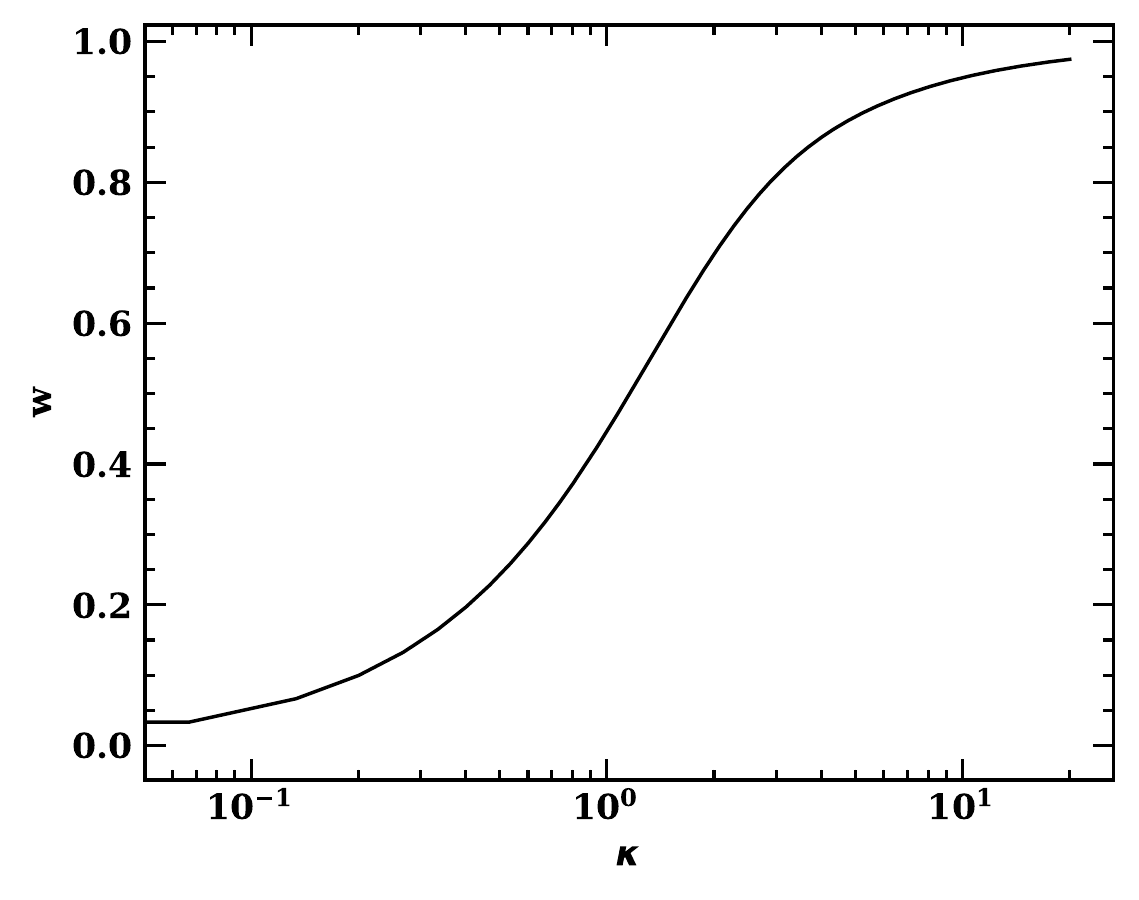}
\caption{Modulation factor or optimal event weight as a function of von Mises concentration parameter $\kappa$. Th function is monotonic, higher concentration parameters are always weighted more. }
\label{fig:weight}       
\end{figure}

\subsubsection{Deep ensembles}
A DNN approach to track reconstruction using deep ensembles gives accurate estimates of individual VM uncertainties $\kappa_i$, \S\ref{sec:perform}. If measurement uncertainties are VM, individual modulation factors are given by eq.~\ref{eqn:weight_func}. 
Then, for a set of deep ensemble predicted emission angles and uncertainties, $\{\hat{\theta}_i, \hat{\kappa}_i\}^N_{i=1}$, weighted maximum likelihood estimators for the polarization parameters eq.~\ref{eqn:qestww},\ref{eqn:uestww} should use
\begin{equation}
\label{eqn:modw}
    w_i = \mu_i = \frac{I_1(\hat{\kappa}_i)}{I_0(\hat{\kappa}_i)}.
\end{equation}
Fig.~\ref{fig:weight} plots the optimal weighting (or equivalently, the modulation factor) as a function of $\kappa$. As expected, $0 \leq \mu_i \leq 1$ and an increasing concentration parameter $\kappa$ (decreasing uncertainty) always increases the event weight.

\subsection{Performance}
\label{sec:perform2}
It is now possible to compare the full polarimetery data analysis pipeline for a DNN approach with the classical moment analysis. Track reconstruction, step 1 of the pipeline, has already been compared in \S\ref{sec:perform}; this section will reveal how both track reconstruction improvements and uncertainty estimation affect polarization recovery. The quality of polarization recovery is measured by the MDP$_{99}$, eq.\ref{eqn:mdpw}. 

\begin{figure}[t]
\sidecaption
\includegraphics[scale=.75]{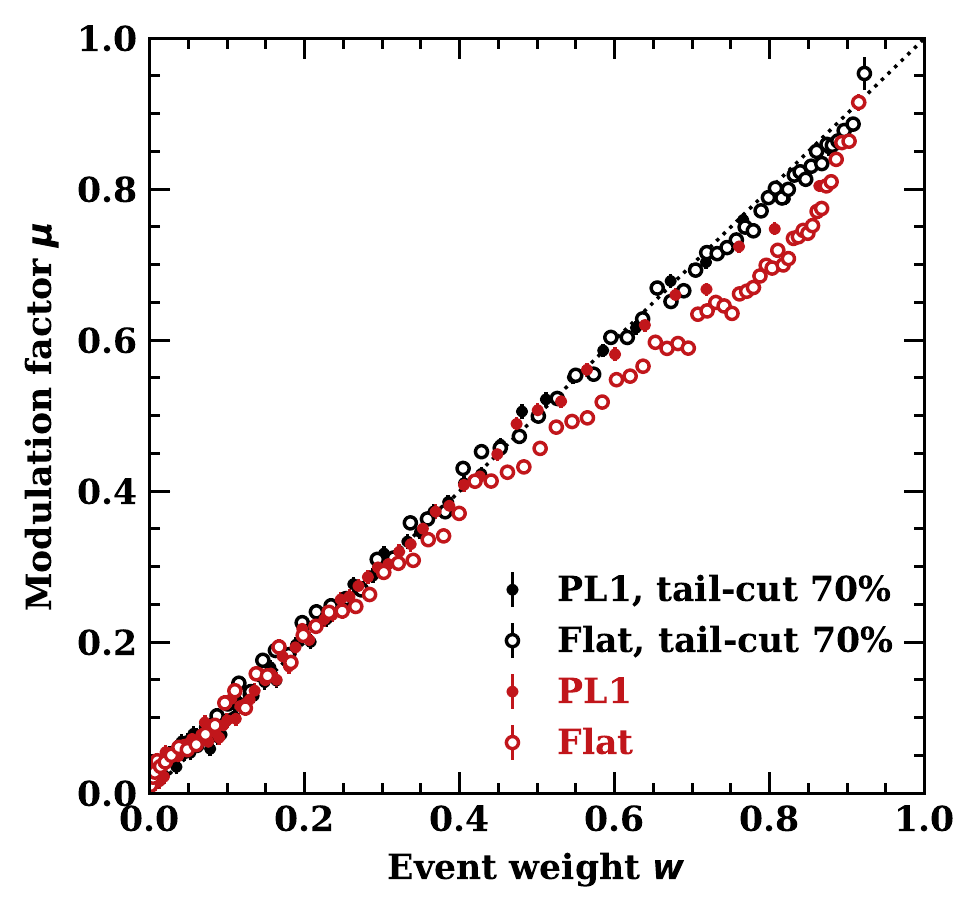}
\caption{Measured $\mu$ as a function of deep ensemble event weight $w$ for large test data sets of $1-10$keV simulated events. Each $\mu$ bin contains 20,000 individual track events. Closed circles represent a PL1 source ($dN/dE = E^{-1}$) convolved with IXPE's effective area, open circles a flat spectrum. Spectra with tail event cuts applied are in black, and full spectra in red.}
\label{fig:weight2}       
\end{figure}

\subsubsection{Weights}
Before comparing methods, it is important to confirm that individual modulation factors $\mu_i$ (or weights $w_i$) predicted by the deep ensemble for each event, eq.\ref{eqn:modw}, are accurate. This can be checked by observing a $100\%$ polarized source and grouping events with same predicted $\mu_i$, then empirically measuring the modulation factor $\mu$ of the group with eq.\ref{eqn:p}. The measured and predicted $\mu$ should always match.

Fig.~\ref{fig:weight2} gives the measured modulation factor $\mu$ as a function of deep ensemble predicted weights $w_i = \mu_i$ for two polarized IXPE datasets with $1.5 \times 10^6$ events.
Any deviation from $y=x$ means the deep ensemble predicted uncertainties are imperfect. When most tail events are removed using a DNN tail cut, \S\ref{sec:removetail}, the deep ensemble predicted weights are close to perfect and show minimal spectral dependence. If tail events are allowed to remain in the spectrum, deep ensemble weights stray from the optimal MLE for highly weighted events. This is because it is difficult for the deep ensemble to learn the appropriate uncertainties $\hat{\kappa}$ for tail events, \S\ref{sec:perform}.
The effect is reduced for the PL1 spectrum where there are fewer high energy events and thus fewer tail events.

\begin{table}[t]
\centering
\begin{tabular}{@{}l l l @{}}
\toprule
{\textbf{Spectrum}}&{\textbf{Method}}&MDP$_{99}$(\%) \\
\midrule
{\textbf{PL2}} & { Mom.} & {5.36 $\pm$ 0.03}\\ 
& { Mom. w/ Ellip. weights} & {4.89 $\pm$ 0.03}\\ 
& { DNN } &{5.10 $\pm$ 0.03 }\\
& { DNN w/ wts.} &{3.99 $\pm$ 0.02}\\
& { DNN w/ wts. 95\%} &{3.98 $\pm$ 0.02 $\leftarrow$}\\
& { DNN w/ wts. 70\%} &{4.08 $\pm$ 0.02}\\
\midrule
{\textbf{PL1}} & { Mom.} & {4.80 $\pm$ 0.02}\\ 
& { Mom. w/ Ellip. weights} & {4.37 $\pm$ 0.02}\\ 
& { DNN } &{4.50 $\pm$ 0.02}\\
& { DNN w/ wts.} & {3.58 $\pm$ 0.01}\\
& { DNN w/ wts. 95\%} &{3.57 $\pm$ 0.01 $\leftarrow$}\\
& { DNN w/ wts. 70\%} &{3.83 $\pm$ 0.01}\\
 \bottomrule
\end{tabular}
\caption{Sensitivity analysis for two power law spectra ($dN/dE \sim E^{-N}$; PL2 for $N=2$, PL1 for $N=1$) each normalized to produce $10^5$ 2-8\,keV photons when folded through IXPE's energy response. ${\rm MDP}_{99}$ gives the sensitivities for the various reconstruction methods; smaller MDP$_{99}$ is better. Mom. denotes moments analysis. DNN denotes deep ensemble. Percentages denote tail cut thresholds, 95\% means events with DNN predicted tail probability greater than 95\% are removed. } 

\label{tab:fom2}
\end{table}

\subsubsection{Comparison}
To expose which parts of the DNN approach are providing the biggest MDP$_{99}$ improvements, table~\ref{tab:fom2} gives the performance breakdown of the different components of the deep ensemble approach. The test datasets are inspired by realistic astrophysical spectra convolved with IXPE's effective area, to give an idea of realized instrument improvements when using DNNs. Since the absolute MDP$_{99}$ values are specific to simulated IXPE events, the relative improvements between methods are the important takeaways as these are likely to generalize to all imaging X-ray polarimeters.

Two baseline classical methods are provided in table~\ref{tab:fom2}, the standard moment analysis described in \S\ref{sec:2} and a weighted version. The weighted version uses event weights derived from moment analysis predicted track ellipticities. Events with higher ellipticities are weighted more in polarization estimation, eq.\ref{eqn:qestww},\ref{eqn:uestww}. The specific function that transforms ellipticities to weights is optimized numerically. The ellipticities provide a rough proxy for emission angle uncertainty. Even this approximate uncertainty calculation can yield significant MDP$_{99}$ improvements.

The base DNN approach predicts emission angles using a deep ensemble but does not use the predicted uncertainty estimates. The improved emission angle estimates yield a marginal improvement over the moment analysis and are worse than a moment analysis with ellipticity based weights. Improvements are small since most astrophysical spectra are soft, containing many low energy photons, where DNN and moment analysis emission angle accuracies are comparable. Clearly, the main improvement to be had in X-ray polarimetry is properly accounting for emission angle heteroskedasticity by appropriately weighting events. This is born out in the weighted deep ensemble MDP$_{99}$ results, where a $< 0.75$ reduction in MDP$_{99}$ from the standard moment analysis is achieved for both spectra. Compared to the classical ellipticity weight approach, the deep ensemble achieves a $\sim 0.82$ MDP$_{99}$.

\begin{svgraybox}
A $0.75$ reduction in MDP$_{99}$ means a $0.75^2 = 0.563$ reduction in the number of counts (and observing time) required to reach the same signal-to-noise ratio. This makes a lot more science possible for an X-ray polarimetry mission, with no changes to the existing hardware.
\end{svgraybox}

Removing tail events using the tail vs peak classifier, \S\ref{sec:removetail}, can provide a small additional sensitivity boost. However, improvements in energy resolution, \S\ref{sec:perform}, require substantially stronger tail exclusion with a threshold of $70\%$ or less. There is a trade-off between achieving good spectral performance and minimizing the MDP$_{99}$ because the DNN tail vs peak classifier is not perfect. When tail event cuts are applied, some peak events are lost. What tail cut threshold should be chosen will depend on the specific telescope and science goals.

\begin{svgraybox}
It is important to note again that all of the results presented in this chapter are for \textit{simulated} IXPE events. Fortunately, tests of the DNN techniques presented here on real IXPE GPD data have yielded similar relative improvements over classical techniques, once real detector systematics are taken into account. If the simulated detector events are close enough to real events, DNN performance will generalize.
\end{svgraybox}

\newpage
\section{Conclusion and future directions} 
\label{sec:6}
X-ray polarimetry and its data analysis is a recently developed and rapidly moving field. Uncertainty aware DNN based approaches yield state-of-the-art signal recovery for imaging X-ray polarimeters and can reduce the required exposure time for telescopes by up to $45\%$.  
Nevertheless, the techniques and discussion presented here should not be considered the final say, but
should form a solid base from which to develop better data analysis for ever-improving X-ray polarimeters. Indeed, this chapter mentions multiple promising areas for improvement. These include:
\begin{itemize}
    \item \textit{A unified approach to tail vs peak event classification.} Training a separate DNN to classify tail events and then cutting events based on its predictions is workable, but clearly not the best solution. A unified DNN approach minimizing a likelihood function that includes the effect of tail probability on emission angle and energy reconstruction should be the ultimate aim. In the case of IXPE, tail events originate from conversions in both the GEM and Beryllium window. These two types of tail event have different morphological properties, so it may also be useful to treat these as separate classes.
    \item \textit{Geometric systematics.} Another shortcut in the DNN approach presented is using square ResNets on multiple rotations of non-square track images, \S\ref{sec:geom}. Again, this works well but is computationally inefficient during evaluation (the DNNs see six images of a single track event). A DNN model that works natively with hexagonal grids should be used for the best results. This DNN model will have the appropriate inductive biases for hexagonal geometries. A promising potential candidate would be the new state-of-the-art visual transformers \cite{dosovitskiy_image_2021}. These DNNs work with coordinates directly, so can be used for any kind of pixel grid. 
\end{itemize}
Deep learning is also a rapidly moving field, so the specific models presented in this chapter may quickly become obsolete. However, any future DNN based data analysis will likely retain the essential ideas: track image processing, individual emission angle uncertainty estimates, full likelihood maximization, and a reduction of events external to the main detector. The author hopes that the next generation of imaging X-ray polarimeters will learn from and improve on the approaches outlined here.

\begin{acknowledgement}
The author would like to thank Roger Romani for a careful reading of the text and the section editors for the opportunity to write this chapter.
\end{acknowledgement}
%


\bibliographystyle{spphys}
\bibliography{references.bib}

\end{document}